\documentclass[aps]{revtex4}
\usepackage{graphicx}
\usepackage{epsfig}
\usepackage{amsmath}
\usepackage{graphics}
\usepackage{latexsym}
\usepackage{subfigure}
\usepackage{xcolor}

\begin{document}           

\title{When does Wenzel's extension of Young's equation for the contact angle of droplets apply? A density functional study}
\author{Sergei A. Egorov$^{1a}$ and Kurt Binder$^2$}

\affiliation{$^1$ Department of Chemistry, University of Virginia,
  Charlottesville, VA 22901, USA\\
  $^{1a}$ Author to whom correspondence should be addressed: sae6z@virginia.edu\\
  $^2$ Institut f\"ur Physik, Johannes Gutenberg Universit\"at Mainz, 55099 Mainz, 
Germany}

\begin{abstract}
  The contact angle of a liquid droplet on a surface under partial wetting conditions differs for a nanoscopically 
rough or periodically corrugated surface from its value for a perfectly flat surface. Wenzel's relation attributes 
this difference simply to the geometric magnification of the surface area (by a factor $r_{\rm w}$), but the validity of
this idea is controversial. We elucidate this problem by model calculations for a sinusoidal corrugation of the
form $z_{\rm wall}(y) = \Delta\cos(2\pi y/\lambda)$ , for a potential of short range $\sigma_{\rm w}$ acting from the wall on the
fluid particles. When the vapor phase is an ideal gas, the change of the wall-vapor surface tension can be
computed exactly, and corrections to Wenzel's equation are typically of order
$\sigma_{\rm w}\Delta/\lambda^2$. For fixed
$r_{\rm w}$ and fixed $\sigma_{\rm w}$ the approach to Wenzel's result with increasing $\lambda$ may be nonmonotonic  and this
limit often is only reached for $\lambda/\sigma_{\rm w}>30$. 
For a non-additive binary mixture, density functional theory
is used to work out the density profiles of both coexisting phases both for planar and corrugated walls, as well
as the corresponding surface tensions. Again, deviations from Wenzel's results of similar magnitude as in the
above ideal gas case are predicted. Finally, a crudely simplified description based on the interface Hamiltonian
concept is used to interpret corresponding simulation results along similar lines. Wenzel's approach is found to
generally hold when $\lambda/\sigma_{\rm w}\gg 1$, $\Delta/\lambda<1$, and conditions avoiding proximity of wetting or filling transitions.
\end{abstract}

\maketitle

\section{Introduction}
\label{section1}

\begin{figure}
\includegraphics[scale=0.5]{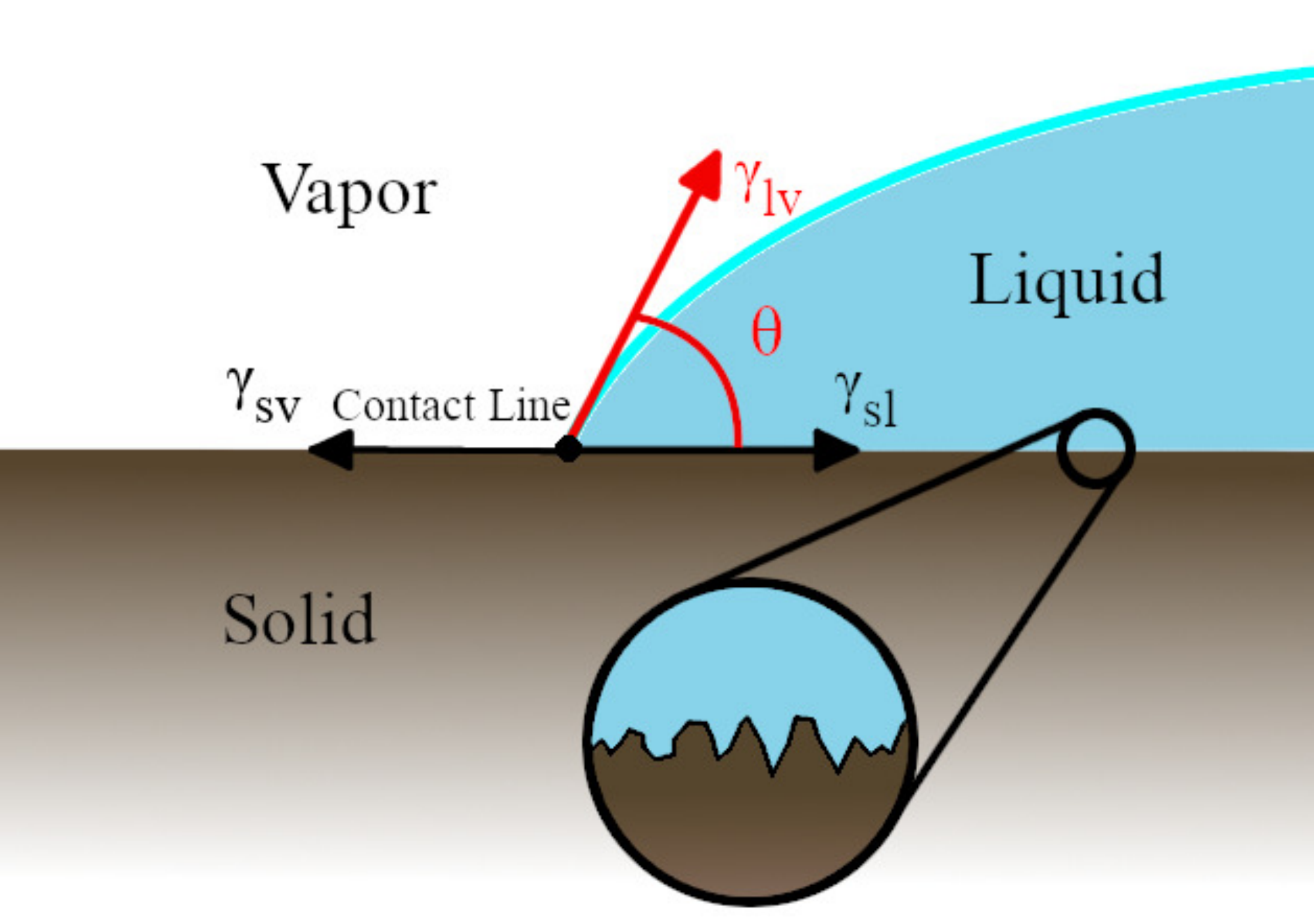} 
\caption{Contact angle $\theta$ of a stable liquid droplet on a solid surface illustrating the force balance on the contact line (the latter is indicated by a black dot). While on a macroscopic level the solid surface is flat, a magnified view often shows that the surface is rough on mesoscopic scales, and it is assumed that due to this roughness the actual surface area of the solid is enhanced by the Wenzel factor $r_{\rm w}$.}  
\label{fig1}
\end{figure}

Wetting of liquids at solid surfaces and related phenomena (spreading of droplets, etc.) are widespread in nature and technology: heterogeneous nucleation of water droplets on dust particles in the atmosphere is important for cloud formation; plants control water droplet motion by special nanopatterns on their leaves; modern technologies such as three-dimensional printing, tissue engineering, formation of templates in microelectronics are just a few examples of industrial applications~\cite{degennes04,butt03,ondarcuhu13,bonn09,erbil14} The basic concept describing a droplet in equilibrium under partial wetting conditions on a planar substrate surface was developed by Young~\cite{young1805} more than 200 years ago. Young's equation expresses the contact angle $\theta$ in terms of the force balance at the contact line (see Fig.~\ref{fig1}):
\begin{equation}
\gamma_{sv}-\gamma_{sl}=\gamma_{lv}\cos\theta,
  \label{young}
\end{equation}
where $\gamma_{sv}$, $\gamma_{sl}$, and $\gamma_{lv}$ are the interfacial tensions between the solid and the vapor phase, between the solid and the liquid phase, and between liquid and vapor phases. It is assumed that the conditions are chosen such that liquid and vapor phases can coexist in thermal equilibrium, and the droplet is almost macroscopically large (so that the excess free energy associated with the three-phase contact line, the line tension~\cite{gibbs61,rowlinson82,amirfazli04,schimmele07}, can be neglected).

Now a surface that looks flat to the naked eye often exhibits considerable roughness on mesoscopic scales 
(see Fig.~\ref{fig1}). Wenzel has suggested that one should modify Eq.~(\ref{young}), taking into account that roughness enhances the actual surface area  $A_r$ of the solid substrate
(relative to the area $A_0$ of a perfectly planar structureless surface) by a factor $r_{\rm w}$, leading to the result that the contact angle gets modified to $\theta^{\ast}$, with~\cite{wenzel36}
\begin{equation}
\cos\theta^{\ast}=r_{\rm w}\cos\theta, \quad r_{\rm w}=A_r/A_0,
  \label{wenzel}
\end{equation}
where we adopt the convention that $\gamma_{sv}$ and $\gamma_{sl}$ are the surface tensions referring to a perfectly planar flat surface of the considered solid substrate. 

\begin{figure}
\includegraphics[scale=0.5]{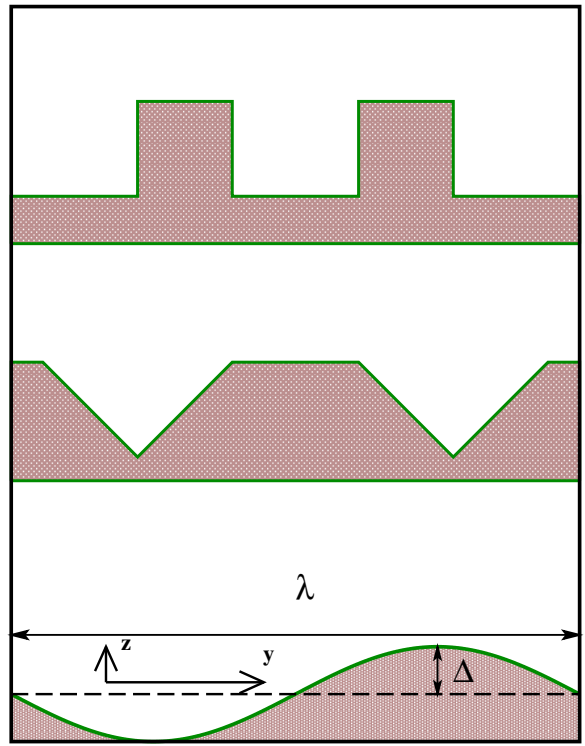} 
\caption{Examples of nanosculpted surfaces with regular periodic roughness: pillars (top part), grooves (middle part), and sinusoidal periodic variation of the surface height $z(x,y)=\Delta\cos(2\pi y/\lambda)$ (bottom part), where both $\Delta$ and $\lambda$ are large in comparison with atomic diameters but small in comparison with the radius of a droplet, as considered in Fig.~\ref{fig1}. For simplicity, only inhomogeneity in one spatial direction ($y$) is assumed.}  
\label{fig2}
\end{figure}

Now roughness as sketched in the lower part of Fig.~\ref{fig1} implies quenched (i.e. frozen-in) disorder at the surface, and such quenched disorder is clearly always a major stumbling block to microscopic understanding in terms of statistical mechanics~\cite{binder11b,li90}. Already the prediction of the suitably averaged surface tensions
$[\gamma_{sv}]_{\rm av}$, $[\gamma_{sl}]_{\rm av}$  (where $[\cdots]_{\rm av}$ denotes averaging over the distribution of the random surface structure) is a nontrivial problem. Moreover, this roughness is expected to affect significantly the dynamics of moving contact lines: the latter experience a rugged free energy landscape, the contact line may get pinned at local minima of this free energy, and thermal activation may be needed to overcome free energy barriers hindering contact line
motions~\cite{erbil14,johnson64,dettre64,swain98,wolansky98,quere08}. 
In fact, contact angle hysteresis (i.e. significant differences between the contact angle of advancing and receding droplet) is a major source of ambiguity in the interpretation of experimental measurements of contact angles~\cite{erbil14,velarde11}. Moreover, when we imagine a Fourier decomposition of the local height $z=h(x,y)$ of a rough surface (relative to the corresponding ideal planar surface $z=0$), we expect that a broad spectrum of wavelengths will contribute, but intuitively it is plausible that Eq.~(\ref{wenzel})  should not include roughness on the scale of a few atoms~\cite{swain98,wolansky98,quere08}. The latter problem as well as the problem of how to average over the disorder~\cite{swain98} is avoided when one considers regular rather than random roughness (Fig.~\ref{fig2}). Such topographically structured surfaces,  with periodically arranged grooves or pillars constitute a very active topic of research, both experimentally (e.g.~\cite{herminghaus08,hofmann10,butt13,xu14}) and from the point of view of theory  (e.g.~\cite{tretyakov16,rascon00,mickel11,berim11,malijevsky14,malijevsky14b,svoboda15,zhou18,malijevsky19})
and simulation 
(e.g.~\cite{xu14,daub10,grzelak10,leroy11,kumar13,chialvo13,tretyakov13,fortini13,svoboda15,ambrosia18}).
Despite this large effort, a clear picture concerning the validity of Wenzel's equation has not yet emerged: in most cases it was found not to hold, at least for the conditions studied; in a few cases it even predicted a qualitatively wrong trend. So the controversy~\cite{erbil14} raised by provocative criticisms~\cite{gao07} about basic failures of Wenzel's approach remains unresolved.  

A crucial clue for understanding why it is difficult to resolve this issue is the well-known fact that the solid-vapor interfacial tension $\gamma_{sv}$, as well as the solid-liquid interfacial tension $\gamma_{sl}$ are not accessible to direct experimental measurement~\cite{marmur11}, irrespective of whether the solid surface is rough or smooth. So the wettability of solid surfaces, which is a property of great interest for various practical applications, is only inferred indirectly from observations of the contact angle of droplets that are put on these surfaces. Of course, when the droplets are small, the contact angle is expected to depend on the droplet radius $R$, and this is in fact observed in simulations of idealized models (e.g.~\cite{egorov18d}). Even for spherical liquid droplets coexisting with  (slightly supersaturated) vapor in the bulk, the liquid-vapor interface tension $\gamma_{lv}(R)$ depends on the droplet radius $R$~\cite{tolman49}, but in quantitative detail this is still not well understood (see e.g.~\cite{troster18}). Since the line tension~\cite{gibbs61,rowlinson82,amirfazli04,schimmele07} provides another correction to the contact angle (see e.g.~\cite{egorov18d}), interpretation of contact angle measurements may be difficult. In the framework of theoretical models, however,  $\gamma_{sv}-\gamma_{sl}$ (for a perfect ideal planar surface), as well as
$\gamma_{sv}^{\rm rough}-\gamma_{sl}^{\rm rough}$ (for a chosen regular roughness type, such as shown in Fig.~\ref{fig2}) are both directly accessible, and hence the relation
\begin{equation}
\gamma_{sv}^{\rm rough}-\gamma_{sl}^{\rm rough}=r_{\rm w}(\gamma_{sv}-\gamma_{sl}), \quad r_{\rm w}=A_r/A_0,
  \label{wenzel2}
\end{equation}
on which Eq.~(\ref{wenzel}) is based is amenable to a direct test, and the nature of corrections to Eq.~(\ref{wenzel2}) can be elucidated. In particular, it is of interest to study how corrections to 
Eq.~(\ref{wenzel2})  depend on the parameters of the roughness pattern, such as wavelength $\lambda$ and amplitude $\Delta$, in the case of sinusoidal corrugation. A similar approach to this problem was chosen by Grzelak and Errington~\cite{grzelak10} and Fortini and Schmidt~\cite{fortini13}, in the framework of Monte Carlo simulations of atomistic models. Due to finite size effects and statistical errors, only somewhat limited conclusions could be drawn from this work, although the general trend was that Eq.~(\ref{wenzel2}) is useful for values of $\lambda$ that are large relative to the fluid particle size.

In order to avoid misunderstandings, we emphasize that the quantities $\gamma_{sv}$, $\gamma_{sl}$,  
$\gamma_{sv}^{\rm rough}$, and $\gamma_{sl}^{\rm rough}$ are defined, as is standard in statistical thermodynamics, as excess free energies of the vapor ($v$) or liquid ($l$) phases that are caused by the contact with the surface of a solid, which can be thought of as inert rigid ``spectator phase'', providing essentially an external potential acting on the particles of the vapor or liquid, respectively. Of course, the actual properties of this potential depend on whether the surface is perfectly smooth or rough, and correspondingly the excess free energies of the vapor or liquid depend on this roughness.  Therefore the argument that can be occasionally found in the literature that Wenzel's equation (Eq.~(\ref{wenzel}) or (\ref{wenzel2})) is invalid because ``the solid molecules at the interface are not mobile and solid surfaces cannot spontaneously contract to minimize their surface
area''~\cite{erbil14} is clearly an irrelevant and misleading argument.

In the present work, we shall follow Refs.~\cite{grzelak10,fortini13} in considering a sinusoidal corrugation of the surface, but avoid approximating it by an atomistic model for the solid substrate. Thus, in our work both parameters $\lambda$ and $\Delta$ can be continuously varied. Given the fact that fluctuation phenomena (such as critical fluctuations, capillary waves at interfaces etc) play no role when we consider conditions far away from critical points in the bulk and second-order wetting, the method of choice is density functional theory (DFT): the mean-field approximation implied in DFT does not produce any dramatic errors here, and a much wider variation of parameters is possible in comparison with computer simulation methods, ``statistical errors'' not being a problem at all. We deliberately restrict attention to the sinusoidal corrugation in Fig.~\ref{fig2} and shall not explicitly discuss the geometries with grooves and pillars: the sharp edges present in these cases lead to the further complication of an additional line tension $\tau_e$ associated with each edge. This line tension is an excess contribution to the free energies of the homogeneous liquid and vapor phases exposed to such geometries~\cite{henderson05}, and should not be confused with the line tension associated with a three-phase vapor-liquid-surface contact line.
We only focus on conditions of partial wetting of the vapor phase, both
for the planar surface and the sinusoidally corrugated surface;  we
are neither concerned with the wetting or drying transitions, nor with
the possibility that filling of wedges or partial filling of the grooves
formed by the sinusoidal corrugation occurs~\cite{rodriguezrivas15}. 
Our results address the limit
of macroscopically large droplets (droplet radius $R\rightarrow \infty$,
in comparison with all lengths discussed here), and hence there is no
distinction between results in the grand-canonical and canonical
ensembles; since the diameter of the droplet circular baseline then tends
to infinity also, there is no dependence expected on the precise
coordinate of the the center of mass position of the droplet relative to
the structure of the corrugation (so the limit we consider is exactly
opposite to the case where the droplet radius is small in comparison
with the wavelength $\lambda$, where a dependence of droplet properties on
the coordinate $y$ of the droplet center of mass when put on the profile
in Fig.~\ref{fig2} indeed can be expected).
 
As a preliminary step, we consider in Sec.~\ref{subsection2.1} the surface tension $\gamma_{sv}^{\rm rough}$ for an ideal gas exposed to a corrugated wall at which a simple square shoulder potential acts on the gas particles, and derive explicit relations for the corrections to the Wenzel relation, using the notion of ``parallel curves'' (also designated in the literature as ``offset curves'')~\cite{yates74,farouki90}. The only other length scale in this problem then is the range $\sigma_{\rm w}$ of the shoulder potential, and we shall show that the correction to $r_{\rm w}$ is of order $\sigma_{\rm w}\Delta/\lambda^2$. When the wall-fluid potential contains both a repulsive and an attractive part, e.g. for a two-Gaussian potential~\cite{alejandre07} considered in Sec.~\ref{subsection2.2},
even a nonmonotonic variation of the correction to $r_{\rm w}$ will be demonstrated, already for an ideal gas. 

Ideal gas, of course, does not have a liquid-vapor phase transition, and so one cannot test the validity of Wenzel's relation in the form of Eq.~(\ref{wenzel}) for this simple model. In order to go beyond this limitation, in Sec.~\ref{section3} we consider a ``penetrable-sphere'' fluid model.
While such potentials as used here are not realistic when fluids of small molecules are concerned,
they are useful both as generic models to illustrate general features, and for the description of fluids
containing soft colloidal particles (e.g., nanoparticles coated with polymer brushes, etc.)~\cite{lekkerkerker11}. 
In this case, analytic treatment of the interfacial tension is no longer possible, and the corresponding values of $\gamma_{sv}$, $\gamma_{sl}$, and $\gamma_{lv}$ are obtained from the DFT calculations. The corrugation of the surface (along one particular direction) requires performing DFT calculations in 2 dimensions, which imposes certain numerical constraints on the sizes of the systems that can be studied. As a result, in this case the approach to the Wenzel limit cannot be studied at the same level of detail as in the case of ideal gas considered in Sec.~\ref{section2}. As an alternative approach, which does not suffer from the numerical restrictions imposed on DFT (but is less microscopic in its nature), we discuss in the Appendix an interface Hamiltonian~\cite{swain98} treatment of wetting on sinusoidally corrugated surfaces. This method allows one to investigate the approach to Wenzel limit for $\Delta/\lambda \ll 1$ 
in analytical fashion, and also makes it possible to fit the simulation data of Ref.~\cite{grzelak10}.

\section{Ideal Gas Results} 
\label{section2}

In this section we consider ideal gas in contact with a wall having a cosinusoidal corrugation profile (in one direction only) and compute the wall-gas interfacial tension $\gamma_{sv}$ for 2 types of the wall-gas interaction potentials. In both cases, the deviations from the Wenzel's equation are clearly demonstrated,
their origin can be explained, and the approach to the Wenzel limit is discussed. 

\subsection{Square-Shoulder Potential}
\label{subsection2.1}

We consider ideal gas at a wall whose corrugation along the $y$-direction is given by a cosine function with 
the wavelength $\lambda$ and the amplitude $\Delta$, i.e. the wall position 
$z_{\rm wall}$ relative to the planar reference wall at $z=0$ is given by:
\begin{equation}
  z_{\rm wall}(y)=\Delta\cos(2\pi y/\lambda).
  \label{zwally}
\end{equation}
For this case, the Wenzel factor $r_{\rm w}$ is given by:
\begin{equation}
  r_{\rm w}=\frac{1}{\lambda}\int_{0}^{\lambda} dy \sqrt{1+
    \left(\frac{2\pi\Delta}{\lambda}\right)^2\sin^2\left(\frac{2\pi y}{\lambda}\right)}.
  \label{rwcosinus}
\end{equation}

The ideal gas particles interact with the wall with a simple square-shoulder-type potential, which is infinite when the closest distance of the particle to the wall is less than or equal to $\sigma_{\rm w}$ and is zero otherwise. In this case, the ideal gas density at the wall $\rho(y,z)$ has a step-like profile, with the step located at the ``parallel'' (also called ``offset'') curve $z_o(y)$ whose points are located at the distance $\sigma_{\rm w}$ along the corresponding normals to the original ``generator'' curve $z_g(y)\equiv z_{\rm wall}(y)$:
\begin{equation}
\rho(y,z)=0 \quad {\rm for } \quad z_g(y)\le z\le z_o(y);\quad \rho(y,z)=\rho_b \quad {\rm for} \quad z>z_o(y),
  \label{rhozy}
\end{equation}  
where $\rho_b$ is the bulk density of the ideal gas 
(strictly speaking, this result applies only to the case of a non-degenerate offset curve, the precise definition of which is given below, while for the degenerate case the situation is more subtle and will be discussed later).
Clearly, for the case of the perfectly planar flat wall ($\Delta=0$, $z_g(y)=0$, $z_o(y)=\sigma_{\rm w}$),
the density profile does not vary in $y$-direction and one simply has
$\rho(z)=0$ for $0\le z\le \sigma_{\rm w}$ and 
$\rho(z)=\rho_b$ for $z>\sigma_{\rm w}$.

With the above form of the density profiles for the rough and flat surfaces, the ratio of the corresponding gas-wall interfacial tensions is given by:~\cite{sitta16}
\begin{equation}
  \frac{\gamma_{sv}^{\rm rough}}{\gamma_{sv}}=\frac{V_{\rm excl}^{\rm rough}}{V_{\rm excl}},
  \label{gammasvratio}
\end{equation}
where $V_{\rm excl}^{\rm rough}$ and $V_{\rm excl}$ are the volumes excluded for the gas particles (due to the square-shoulder wall potential) at the rough and flat surfaces, respectively. On the other hand, Wenzel's relation given by Eq.~(\ref{wenzel2}) predicts that the ratio of these interfacial tensions is equal to
$r_{\rm w}$, with the latter given by Eq.~(\ref{rwcosinus}). Accordingly, in order to test the validity of the Wenzel's prediction one needs to compare the ratio of the volumes $V_{\rm excl}^{\rm rough}/V_{\rm excl}$ to
$r_{\rm w}$. Due to periodicity of the wall profile in the $y$-direction and the absence of the corrugation in the $x$-direction, in calculating the above excluded volumes it is sufficient to consider $y$ ranging over a single wavelength $\lambda$ and a unit range in the $x$-direction, whereby for the perfectly flat wall one simply obtains
$V_{\rm excl}=\lambda\sigma_{\rm w}d_x$ (with $d_x\equiv 1$ and the value of $\lambda$ taken from the corrugated case with which the comparison is made). At the same time, as already mentioned above in connection with Eq.~(\ref{rhozy}), the calculation of $V_{\rm excl}^{\rm rough}$ requires a precise definition of the offset curves, including the distinction between degenerate and non-degenerate cases, and therefore this subject is briefly discussed next. 

The general notion of offset curves was first introduced by Leibnitz in 1692~\cite{yates74}.
In the present discussion we largely follow the terminology, notation and methodology of the article by Farouki and Neff on the analytic properties of offset 
curves~\cite{farouki90} (note that the absence of the corrugation in the $x$-direction allows us to limit the discussion to the case of planar offset curves). In particular, the (planar) generator curve is taken to be parametrized by variable $t$ (in the range between $t_{\rm min}$ and $t_{\rm max}$) as $\vec{r}_g(t)=[y_g(t),z_g(t)]$, and the corresponding (interior) offset curve is defined by
\begin{equation}
\vec{r}_o(t)=\vec{r}_g(t)-\sigma_{\rm w}\vec{n}(t),
 \label{roffset} 
\end{equation}
where $\vec{n}(t)$ is the unit normal to the generator curve at each point between $t_{\rm min}$ and $t_{\rm max}$. 
In the present case, $y_g(t)=t$ and $z_g(t)=\Delta\cos[2\pi t/\lambda]$.

For the offset curve 
$\vec{r}_o(t)=[y_o(t),z_o(t)]$ one gets~\cite{farouki90}:
\begin{equation}
y_o(t)=y_g(t)-\frac{\sigma_{\rm w}z_{g}^{\prime}(t)}{\sqrt{(y_{g}^{\prime}(t))^2+(z_{g}^{\prime}(t))^2}},
 \label{yoffset} 
\end{equation}
and
\begin{equation}
z_o(t)=z_g(t)+\frac{\sigma_{\rm w}y_{g}^{\prime}(t)}{\sqrt{(y_{g}^{\prime}(t))^2+(z_{g}^{\prime}(t))^2}},
 \label{zoffset} 
\end{equation}
which in the present case yields:
\begin{equation}
  y_o(t)=t+\frac{2\sigma_{\rm w}\pi\Delta\sin[2\pi t/\lambda]/\lambda}
  {\sqrt{1+(2\pi\Delta\sin[2\pi t/\lambda]/\lambda)^2}},
 \label{yoffsetcos} 
\end{equation}
and
\begin{equation}
z_o(t)=\Delta\cos[2\pi t/\lambda]+\frac{\sigma_{\rm w}}{\sqrt{1+(2\pi\Delta\sin[2\pi t/\lambda]/\lambda)^2}}.
 \label{zoffsetcos} 
\end{equation}
It is immediately clear that an offset curve to a cosinusoidal curve is {\em not} a cosinusoidal curve (the same is true for most other functional forms of the generator curves, with only a few exceptions, such as circular curves).

Next, one needs to distinguish between degenerate and non-degenerate offset curves~\cite{farouki90}. The offset curve is said to be non-degenerate when the offset distance $\sigma_{\rm w}$ is smaller than the radius of curvature $R_g(t)$ of the generator curve for all values of $t$ considered, while in the opposite case a degenerate offset curve is obtained.
$R_g(t)$ is given by~\cite{farouki90}:
\begin{equation}
  R_g(t)=\frac{((y_{g}^{\prime}(t))^2+(z_{g}^{\prime}(t))^2)^{3/2}}
  {|y_{g}^{\prime}(t)z_{g}^{\prime\prime}(t)-z_{g}^{\prime}(t)y_{g}^{\prime\prime}(t)|},
 \label{rgt} 
\end{equation}
which yields for the corrugation profile given by Eq.~(\ref{zwally}):
\begin{equation}
  R_g(t)=\left(\frac{\lambda}{2\pi}\right)^2\frac{(1+(2\pi\Delta\sin[2\pi t/\lambda]/\lambda)^2)^{3/2}}
  {\Delta \cos[2\pi t/\lambda]}.
 \label{rgtcost} 
\end{equation}
Taking $t_{\rm min}=0$ and $t_{\rm max}=\lambda$,
the smallest value of $R_g(t)$ is given by:
\begin{equation}
  R_{g}^{\rm min}=\frac{\lambda^2}{4\pi^2\Delta}. 
 \label{rgtmin} 
\end{equation}

\begin{figure}
\includegraphics[scale=0.5]{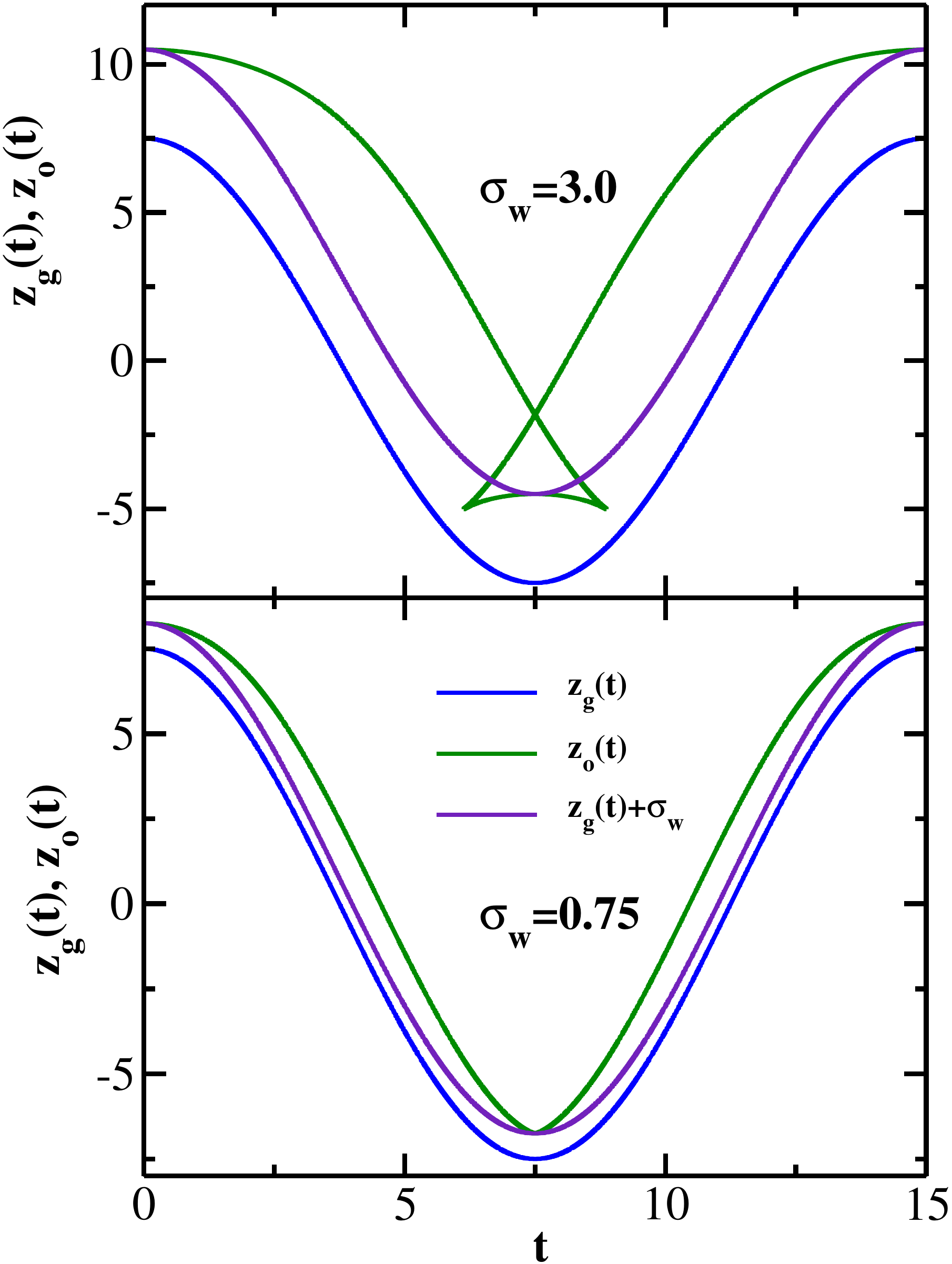} 
\caption{Generator and offset curves as defined in the text for $\lambda=15$ and $\Delta=7.5$,
  also shown is the generator curve displaced vertically by the offset distance $\sigma_{\rm w}$. 
Upper panel: $\sigma_{\rm w}=3.0$; lower panel: $\sigma_{\rm w}=0.75$.}  
\label{fig3}
\end{figure}

We now illustrate the difference between the non-degenerate and degenerate curves by choosing specific values of the wavelength $\lambda$ and the amplitude $\Delta$, calculating $R_{g}^{\rm min}$ from Eq.~(\ref{rgtmin}), and considering the two cases with $\sigma_{\rm w}<R_{g}^{\rm min}$ and $\sigma_{\rm w}>R_{g}^{\rm min}$, respectively.  
Specifically, taking $\lambda/d_x=15$ and $\Delta/d_x=7.5$, one obtains $R_{g}^{\rm min}/d_x=0.76$, and so setting
$\sigma_{\rm w}/d_x=0.75$ would give a non-degenerate offset curve (from here on we report all lengths in units of $d_x$ and no longer write it explicitly). This situation is illustrated in the lower panel of Fig.~\ref{fig3} which shows the generator and offset curves for the above parameter values; also shown is the generator curve simply shifted vertically by $\sigma_{\rm w}$, in order to illustrate the deviation of the offset curve from the cosinusoidal shape of $z_g(t)$ as mentioned earlier. By contrast, setting $\sigma_{\rm w}=3.0$ gives a degenerate off-set curve, as shown in the upper panel of Fig.~\ref{fig5}. Its most characteristic feature is the presence of cusps and self-intersections~\cite{farouki90}.

With the above definitions, we now return to the calculation of the ratio $V_{\rm excl}^{\rm rough}/V_{\rm excl}$ which needs to be compared to $r_{\rm w}$ in order to assess the range of validity of the Wenzel's relation. 
Starting with a simpler non-degenerate case and taking $t_{\rm min}=0$ and $t_{\rm max}=\lambda$,
this ratio is given by~\cite{farouki90}:
\begin{equation}
  \frac{V_{\rm excl}^{\rm rough}}{V_{\rm excl}}=\frac{1}{2\lambda}\int_{0}^{\lambda}
       \left[1+|1-\frac{\sigma_{\rm w}}{R_g(t)}|\right]{\sqrt{(x_{g}^{\prime}(t))^2+(z_{g}^{\prime}(t))^2}}dt.
 \label{ratioroughflat} 
\end{equation}

Note that in the case of macroscopic-scale corrugation (where 
$\sigma_{\rm w}/R_{g}^{\rm min}\rightarrow 0$) one obtains:
\begin{equation}
  \frac{V_{\rm excl}^{\rm rough}}{V_{\rm excl}}=\frac{1}{\lambda}\int_{0}^{\lambda}
       {\sqrt{(x_{g}^{\prime}(t))^2+(z_{g}^{\prime}(t))^2}}dt=r_{\rm w},
 \label{ratioroughflatwenzel} 
\end{equation}
confirming that in the macroscopic limit the Wenzel relation is indeed recovered.
From Eqs.~(\ref{rgtcost}) and~(\ref{rgtmin}) which imply that the order of magnitude of
$R_g(t)$ is $\lambda^2/\Delta$ and
Eq.~(\ref{ratioroughflat}) we already recognize that the
order of magnitude of corrections to the Wenzel equation must in general be
$\sigma_{\rm w}\Delta/\lambda^2$. 

In order to analyze the deviations from Wenzel's relation for microscopic-scale corrugations in more detail, one needs to consider the situation where $\sigma_{\rm w}$ is comparable to or larger than $R_{g}^{\rm min}$, and therefore one needs to deal with both non-degenerate and degenerate offset curves. While in the former case the quantity of interest $V_{\rm excl}^{\rm rough}/V_{\rm excl}$ is straightforwardly given by Eq.~(\ref{ratioroughflat}) (and can be easily evaluated numerically), in the latter case the degenerate offset curve needs to be ``trimmed''~\cite{farouki90}, which means that the triangular-shaped area between the self-intersection point and the two cusp points in the upper panel of Fig.~\ref{fig3} must be removed. The corresponding expression for
the ratio $V_{\rm excl}^{\rm rough}/V_{\rm excl}$ becomes rather more involved, but can still be evaluated numerically in a straightforward manner.

\begin{figure}
\includegraphics[scale=0.5]{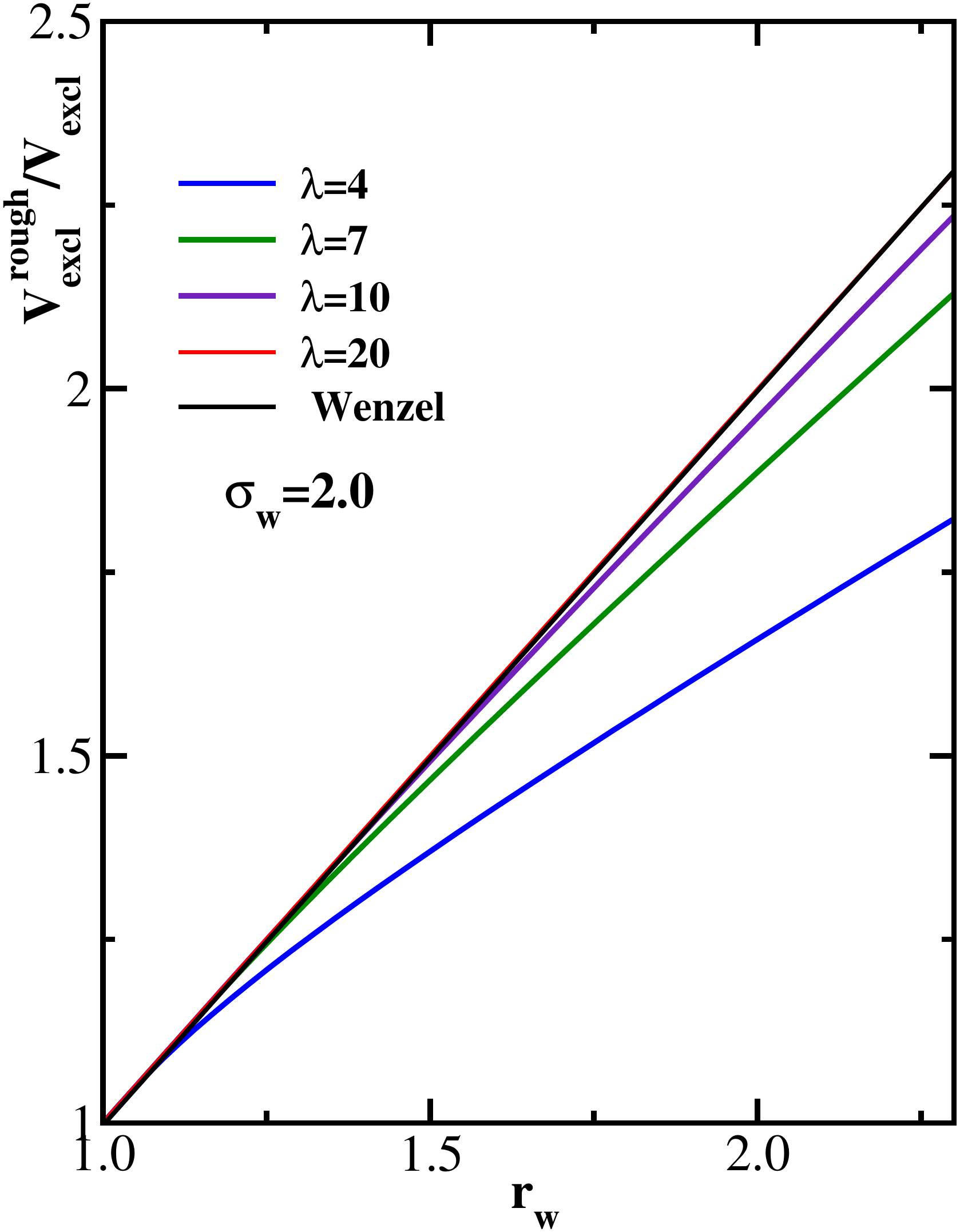} 
\caption{Ratio $V_{\rm excl}^{\rm rough}/V_{\rm excl}$ of corrugated and flat inaccessible volumes vs the Wenzel factor $r_{\rm w}$ for several values of $\lambda$, as indicated; the offset distance is fixed at $\sigma_{\rm w}=2.0$. The case
 $\lambda=20$ is already indistinguishable from Wenzel's result here.} 
\label{fig4}
\end{figure}

\begin{figure}
\includegraphics[scale=0.5]{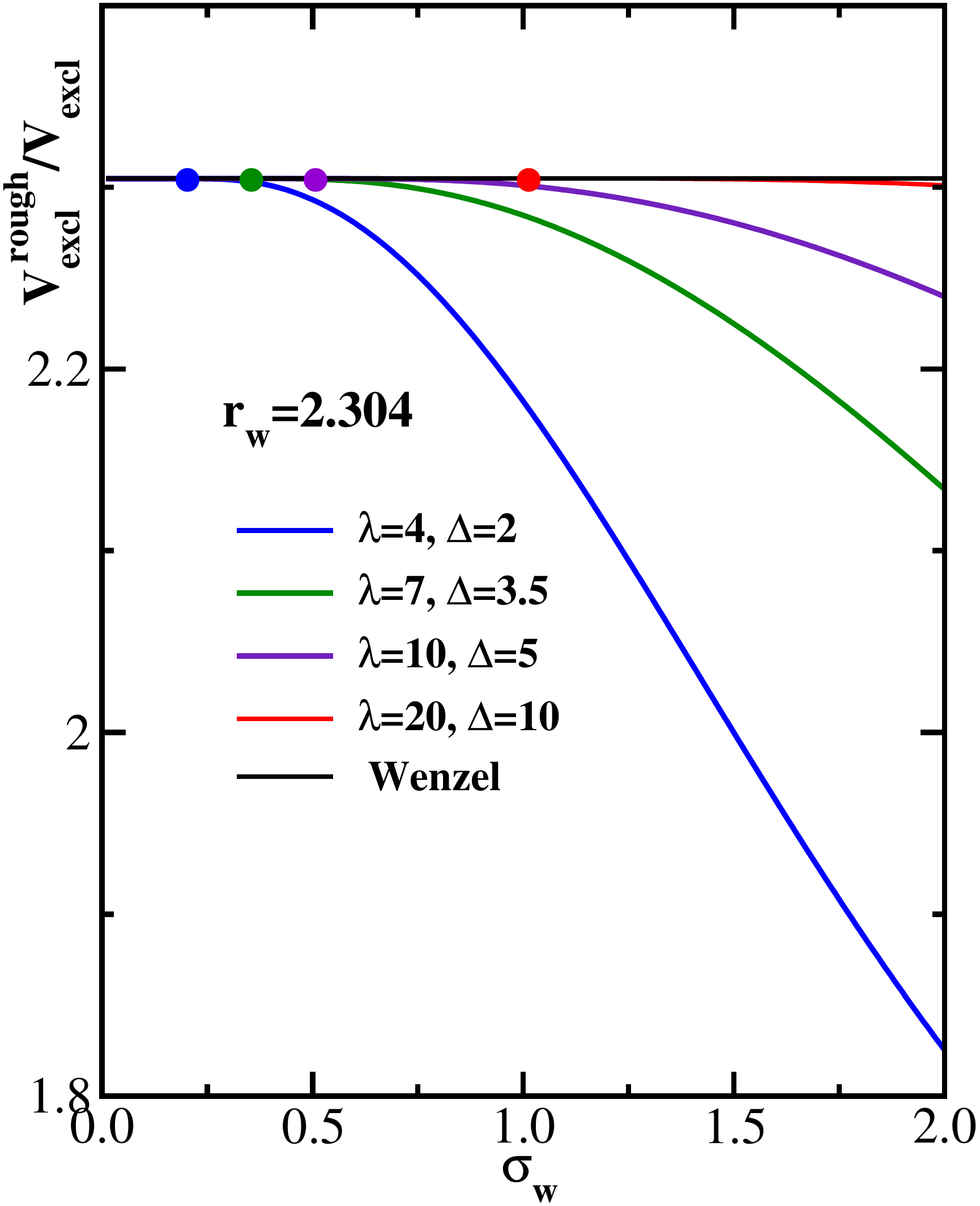} 
\caption{Ratio $V_{\rm excl}^{\rm rough}/V_{\rm excl}$ of corrugated and flat inaccessible volumes vs the offset distance $\sigma_{\rm w}$ for several values of $\lambda$ and $\Delta$, as indicated; the Wenzel ratio is fixed at $r_{\rm w}=2.304$ ($\lambda/\Delta=2.0$). The circles mark the values of the offset distance $\sigma_{\rm w}=R_{g}^{\rm min}=\lambda^2/(4\pi^2\Delta)$, corresponding to the boundary between non-degenerate and degenerate offset curves.} 
\label{fig5}
\end{figure}

Next, we proceed to calculate the ratio $V_{\rm excl}^{\rm rough}/V_{\rm excl}$ as discussed above for several representative values of $\lambda$, $\Delta$, and $\sigma_{\rm w}$.
First, we set $\sigma_{\rm w}=2$, choose several values of the corrugation wavelength $\lambda$, and vary 
the corrugation amplitude $\Delta$. We show the corresponding results for the ratio 
$V_{\rm excl}^{\rm rough}/V_{\rm excl}$ as a function of the Wenzel's ratio $r_{\rm w}$ in Fig.~\ref{fig4} (clearly, Wenzel's relation itself simply gives a straight line with slope 1). One sees that the deviations from Wenzel relation are more pronounced for smaller wavelengths $\lambda$ and increase with increasing degree of corrugation $r_{\rm w}$.

Conversely, one can fix the value of $r_{\rm w}$ and vary the offset distance $\sigma_{\rm w}$. This is done in
Fig.~\ref{fig5}, where we choose several values of the corrugation wavelength $\lambda$, set 
the corrugation amplitude $\Delta=\lambda/2$ (corresponding to $r_{\rm w}=2.304$) and plot the ratio 
$V_{\rm excl}^{\rm rough}/V_{\rm excl}$ as a function of the offset distance $\sigma_{\rm w}$. As expected, the deviations 
from Wenzel's relation increase with increasing $\sigma_{\rm w}$. 
As discussed above, the equality
$\sigma_{\rm w}=R_{g}^{\rm min}=\lambda^2/(4\pi^2\Delta)$ marks the boundary between degenerate and non-degenerate offset curves. We mark the corresponding values of $\sigma_{\rm w}$ as circles in Fig.~\ref{fig5}. Interestingly,
for $\sigma_{\rm w}<R_{g}^{\rm min}$ the Wenzel's relation appears to hold to very good accuracy (although even in this range it cannot be exact, as follows from the comparison of Eqs.~(\ref{ratioroughflat}) and (\ref{ratioroughflatwenzel})).   

\begin{figure}
\includegraphics[scale=0.5]{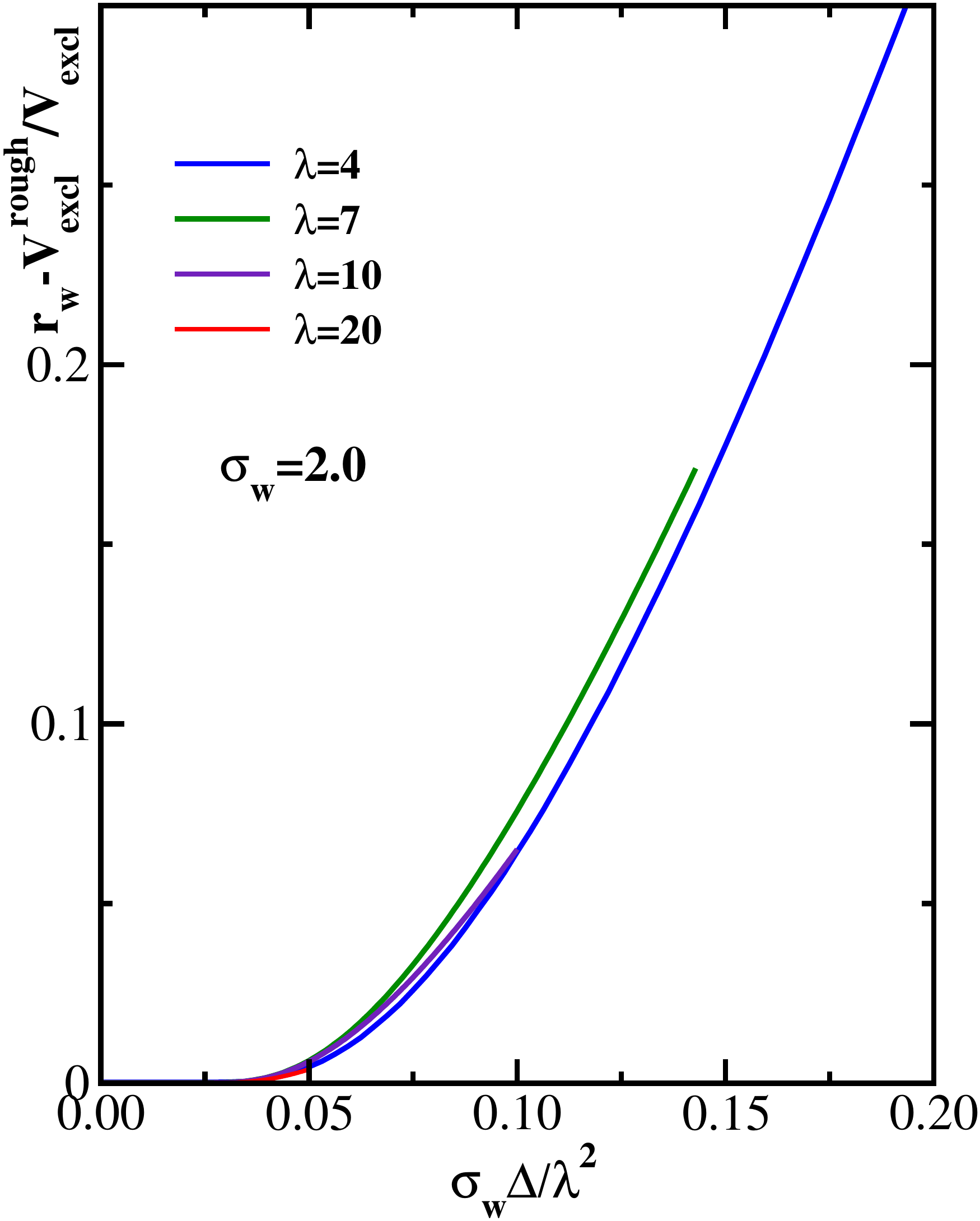} 
\caption{The correction to the Wenzel's relation $r_{\rm w}-V_{\rm excl}^{\rm rough}/V_{\rm excl}$ as a function 
 of dimensionless ratio $\sigma_{\rm w}\Delta/\lambda^2$ for several values of $\lambda$, as indicated; the offset distance is fixed at $\sigma_{\rm w}=2.0$.} 
\label{fig6}
\end{figure}

Returning to Fig.~\ref{fig4}, its inspection suggests again that the correction to the Wenzel's relation, i.e. the difference $r_{\rm w}-V_{\rm excl}^{\rm rough}/V_{\rm excl}$, is of the order of dimensionless ratio
$\sigma_{\rm w}\Delta/\lambda^2$ (which contains all the three relevant length scales in the problem).  
This is further illustrated in Fig.~\ref{fig6}, which shows $r_{\rm w}-V_{\rm excl}^{\rm rough}/V_{\rm excl}$ as a function of 
$\sigma_{\rm w}\Delta/\lambda^2$ for several values of wavelength $\lambda$. Specifically, for each value of $\lambda$ we fix the offset distance at $\sigma_{\rm w}=2.0$ and vary the amplitude $\Delta$ in the range between 0 and $\lambda/2$ in order to generate these results. One sees that the magnitude of the correction decreases with increasing $\lambda$, but for each specific value of $\lambda$ the magnitude of the correction is indeed on the order of $\sigma_{\rm w}\Delta/\lambda^2$. 
 
\subsection{Two-Gaussian Potential}
\label{subsection2.2}

\begin{figure}
\includegraphics[scale=0.5]{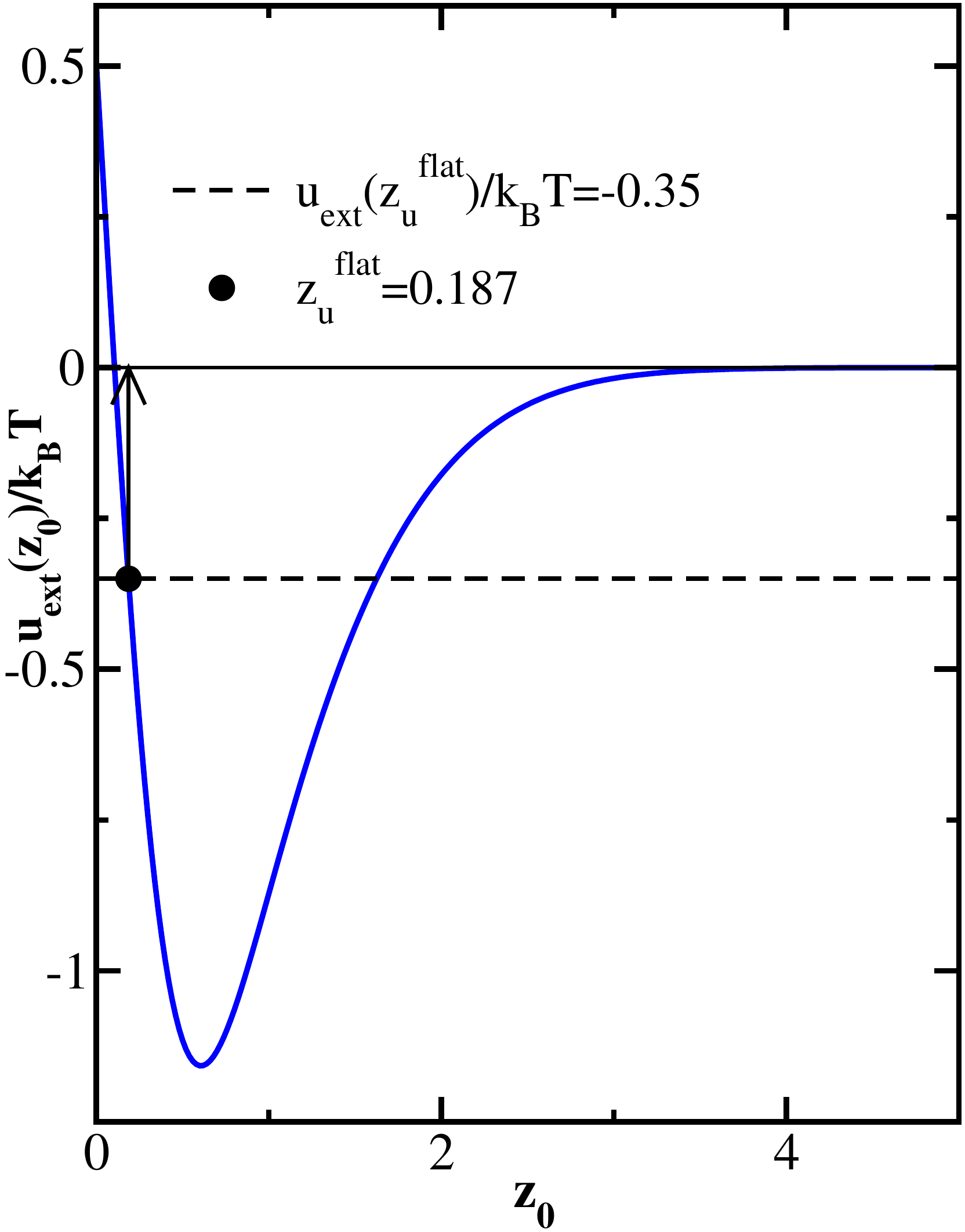} 
\caption{The (dimensionless) external potential (as given by Eq.~(\ref{uextyz})) experienced by the gas particle at a perfectly flat wall as a function of its distance $z_0$ from the wall. The black dot indicates the value
$z_0=z_{u}^{\rm flat}=0.187$ at which $\beta u_{\rm ext}(z_{u}^{\rm flat})=-0.35$, which will be used in constructing Fig.~\ref{fig12} below.} 
\label{fig7}
\end{figure}

The square-shoulder wall-gas potential discussed in Sec.~\ref{subsection2.1} is uniform along the substrate profile, i.e. the offset distance $\sigma_{\rm w}$ is the same along the corrugation direction for all values of $y$. A more realistic model would represent the substrate as a collection of individual particles distributed uniformly with number density $\rho_s$ for all $z(y)\le z_g(y)$, with each substrate particle interacting with a given gas particle via an isotropic pairwise potential $u_{sg}(r)$. In this model, the external potential experienced by an ideal gas particle located at $\vec{r}_0$ is obtained by integrating the pair potential $u_{sg}(|\vec{r}_0-\vec{r}|)$
over all values of $\vec{r}$ corresponding to the space occupied by the substrate particles.
While in the simulation literature it is common to use the familiar Lennard-Jones (LJ) functional form for the potential $u_{sg}(r)$~\cite{grzelak10}, its numerical integration within the present DFT approach can be challenging, and therefore we instead employ a two-Gaussian potential~\cite{alejandre07}, whose general shape is quite similar to the LJ form (except that it stays bounded for small values of $r$, which makes it possible to avoid the numerical difficulties associated with the integration of LJ potential).
In particular, we take the following form:
\begin{equation}
u_{sg}(r)=\epsilon_1\exp[-(\alpha_1r)^2]-\epsilon_2\exp[-(\alpha_2r)^2],
  \label{twogauss}
\end{equation}  
where we follow Ref.~\onlinecite{alejandre07} and take the following values of the well-depth and range parameters: $\epsilon_1=85.2k_BT$, $\epsilon_2=2.13k_BT$, $\alpha_1=2.02/d_x$, and $\alpha_2=0.63/d_x$, where $d_x=1$ is again our unit of length. Setting the (dimensionless) substrate density $\rho_sd_{x}^{3}=1$, we obtain for a gas particle located at $\vec{r}_0=(x_0,y_0,z_0)$ the following external potential
$u_{\rm ext}(y_0,z_0)$ due to its interaction with the substrate atoms:
\begin{equation}
  u_{\rm ext}(y_0,z_0)=\epsilon_1\frac{\sqrt{\pi}}{\alpha_1}
  \int_{-\infty}^{\infty} dy e^{-\alpha_{1}^{2}(y-y_0)^2}
  \int_{-\infty}^{z_g(y)} dz e^{-\alpha_{1}^{2}(z-z_0)^2}
  -\epsilon_2\frac{\sqrt{\pi}}{\alpha_2}
  \int_{-\infty}^{\infty} dy e^{-\alpha_{2}^{2}(y-y_0)^2}
  \int_{-\infty}^{z_g(y)} dz e^{-\alpha_{2}^{2}(z-z_0)^2}.
  \label{uextyz}
\end{equation}
Note that due to the symmetry of our model (the substrate extending infinitely along the $x$ axis without corrugation) the integration over $x$ has been performed analytically, and the resulting integrated gas-substrate potential does not depend on $x_0$. As mentioned above, the pair potential $u_{sg}(r)$ stays bounded for small values of $r$, and in order to prevent the penetration of gas atoms inside the substrate we use the above form for
$u_{\rm ext}(y_0,z_0)$ only for $z_0\ge z_g(y_0)$ (gas atom above the substrate), while for $z_0 < z_g(y_0)$ we set 
$u_{\rm ext}(y_0,z_0)=\infty$ (hard wall). In the case of perfectly planar flat wall ($\Delta=0$, $z_g(y)=0$), the
$y-$integration in Eq.~(\ref{uextyz}) can be performed analytically, and the resulting external potential is a function of $z_0$, it is shown in Fig.~\ref{fig7}. 

The density profile of the ideal gas above the substrate (normalized by the bulk density $\rho_b$) is given by the Boltzmann expression:
\begin{equation}
  \frac{\rho(y_0,z_0)}{\rho_b}=e^{-\beta u_{\rm ext}(y_0,z_0)},
  \label{boltzmann}
\end{equation}
where $\beta=1/(k_BT)$. For numerical convenience, we define a function describing the relative deviation of the density from its bulk value:
\begin{equation}
  R(y_0,z_0)=\frac{\rho(y_0,z_0)}{\rho_b}-1,
  \label{ryz}
\end{equation}
which decays to zero for large distances away from the substrate and therefore can be integrated over the entire space occupied by the ideal gas above the substrate (to facilitate the comparison with the Wenzel relation).

As discussed at the beginning of this Section, the present model differs in one important respect from the square-shoulder model described in Sec.~\ref{subsection2.1} -- namely, in the presence of corrugation, the potential $u_{\rm ext}(y_0,z_0)$ depends not only on the distance of the gas particle from the wall, but also on its location $y_0$ along the profile. In order to illustrate this dependence, we compute the function 
$R_g(t)\equiv R(y_g(t),z_g(t))$ from Eq.~(\ref{ryz}) along the substrate profile $z_g(t)$ from $t_{\rm min}=0$ up to
$t_{\rm max}=\lambda/2$ for several values of wavelength $\lambda$ and Wenzel's ratio $r_{\rm w}$. Note that Wenzel's relation implicitly assumes that $R_g(t)$ is independent of $t$ and is equal to the corresponding value for the flat profile ($r_{\rm w}=1$).

\begin{figure}
\includegraphics[scale=0.5]{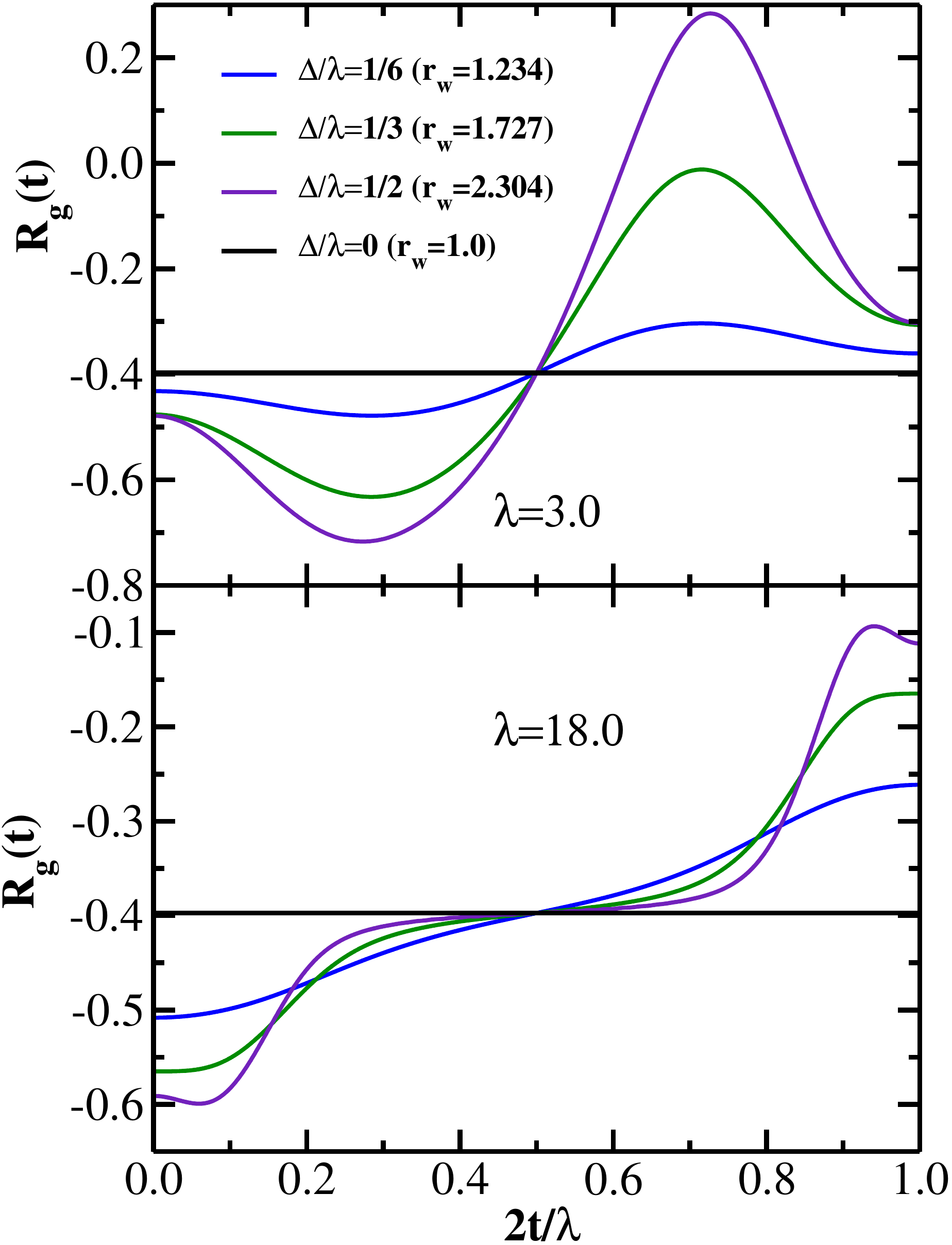} 
\caption{Function $R_g(t)$ describing
the relative deviation of the density from its bulk value along the
contour $y_g(t)$, $z_g(t)$ in the $(y,z)$ plane vs $2t/\lambda$ for several values of the Wenzel's ratio $r_{\rm w}$ as indicated. Upper panel: substrate wavelength $\lambda=3$, lower panel: substrate wavelength $\lambda=18$; black line indicates the result for a flat substrate.} 
\label{fig8}
\end{figure}

In Fig.~\ref{fig8} we plot our numerical results for $R_g(t)$ vs $2t/\lambda$ for several values of the
ratio $\Delta/\lambda$ (each of which corresponds to a particular value of $r_{\rm w}$):  
$\Delta/\lambda=1/6$ ($r_{\rm w}=1.234$), $\Delta/\lambda=1/3$ ($r_{\rm w}=1.727$),
and $\Delta/\lambda=1/2$ ($r_{\rm w}=2.304$); also included is the result for a flat profile
($\Delta=0$, $r_{\rm w}=1$). Upper panel presents the results for $\lambda=3$, and lower panel -- for $\lambda=18$. One immediately observes that in contrast to Wenzel's implicit assumption, $R_g(t)$ does depend on $t$ and deviates from the flat value everywhere except for the midpoint $t=\lambda/4$. This dependence can be rationalized by noting that a gas atom located at the top of the substrate ($t=0$) experiences fewer interactions with nearby substrate atoms as compared to the gas atom located at the bottom of the curve ($t=\lambda/2$); this difference is reflected in the corresponding values of $u_{\rm ext}(y_0,z_0)$, and, therefore, $R_g(t)$. One also notes that the deviation of $R_g(t)$ from the flat result decreases with decreasing corrugation (smaller values of $r_{\rm w}$), as one would expect.

\begin{figure}
\includegraphics[scale=0.5]{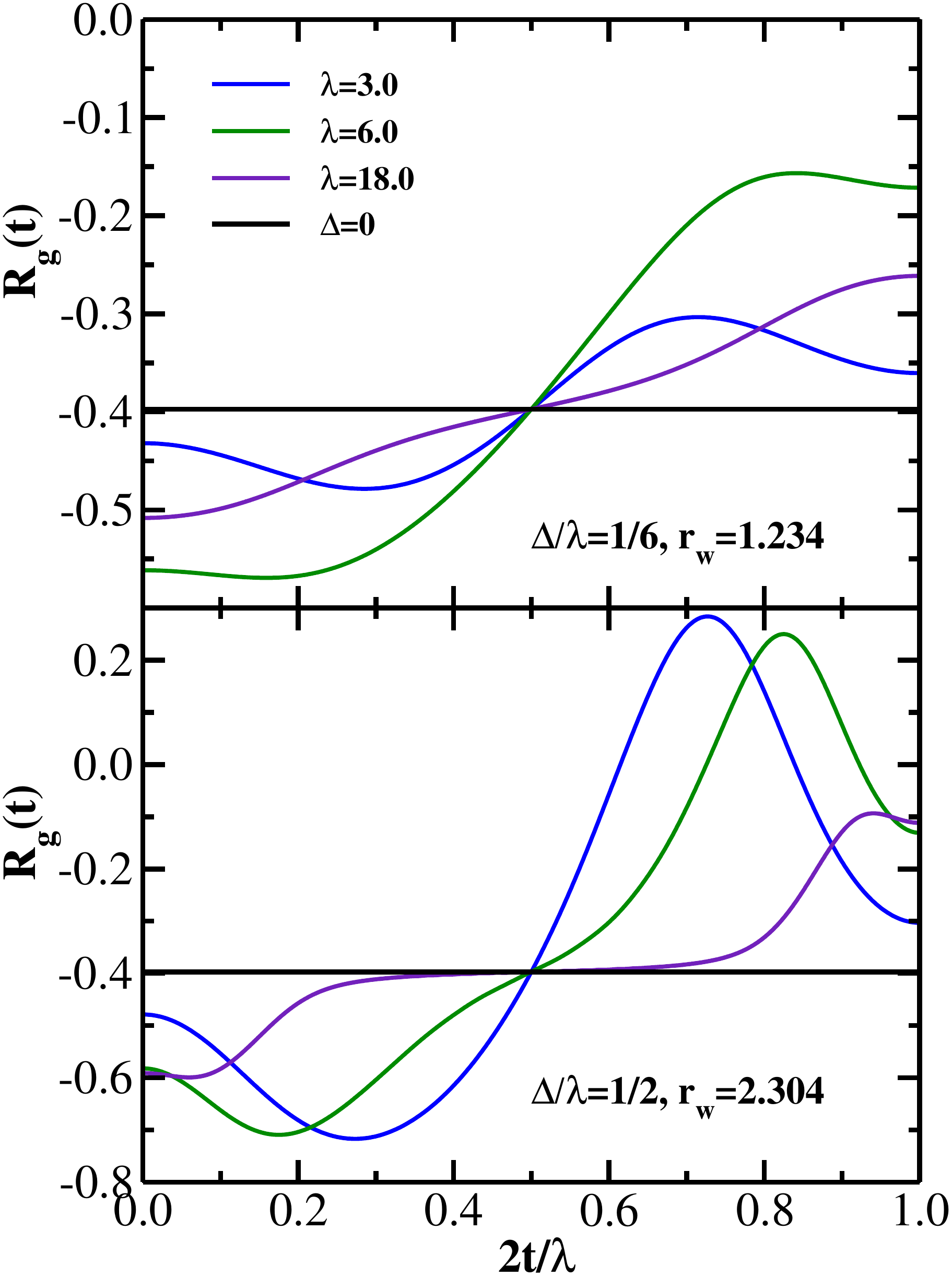} 
\caption{Function $R_g(t)$ describing
the relative deviation of the density from its bulk value along the
contour $y_g(t)$, $z_g(t)$ in the $(y,z)$ plane vs $2t/\lambda$ for several values of the substrate wavelength as indicated. Upper panel: Wenzel's factor $r_{\rm w}=1.234$, lower panel: Wenzel's factor $r_{\rm w}=2.304$; black line indicates the result for a flat substrate.} 
\label{fig9} 
\end{figure}

Next, in Fig.~\ref{fig9} we plot our numerical results for $R_g(t)$ vs $2t/\lambda$ for several values of the substrate wavelength: $\lambda=3$, $\lambda=6$, and $\lambda=18$. Upper panel presents the results for $\Delta/\lambda=1/6$ ($r_{\rm w}=1.234$), and lower panel -- for $\Delta/\lambda=1/2$ ($r_{\rm w}=2.304$); also included in both panels is the result for a flat profile ($\Delta=0$). Interestingly, while the deviation of $R_g(t)$ from the flat result generally does decrease with increasing $\lambda$, it happens non-monotonically, with the deviation for $\lambda=6$ in the upper panel being somewhat larger compared to $\lambda=3$. 

Having considered the behavior of the gas-substrate integrated potential along the substrate profile $z_g(t)$, we next draw a sequence of parallel curves $z_o(t)$ all equally spaced by a small distance $dz=0.01d_x$ from each other. By integrating the function $R(y_o(t),z_o(t))$ (given by Eq.~(\ref{ryz})) along each of these curves from $t_{\rm min}=0$ up to
$t_{\rm max}$ (chosen such that $y_o(t_{\rm max})=\lambda/2$), we obtain a function $\bar{R}(z)$, where $z$ is the distance between a given offset curve $z_o(t)$ and the generator curve $z_g(t)$:
\begin{equation}
\bar{R}(z)=\int_{t_{\rm min}}^{t_{\rm max}}R(y_o(t),z_o(t))dt.
  \label{rz}
\end{equation}  
Note that according to the Wenzel's relation, the function $2\bar{R}(z)/(\lambda r_{\rm w})$ should coincide with the corresponding flat result
$\bar{R}_{\rm flat}(z)$ for all values of the corrugation and wavelength.

\begin{figure} 
\includegraphics[scale=0.5]{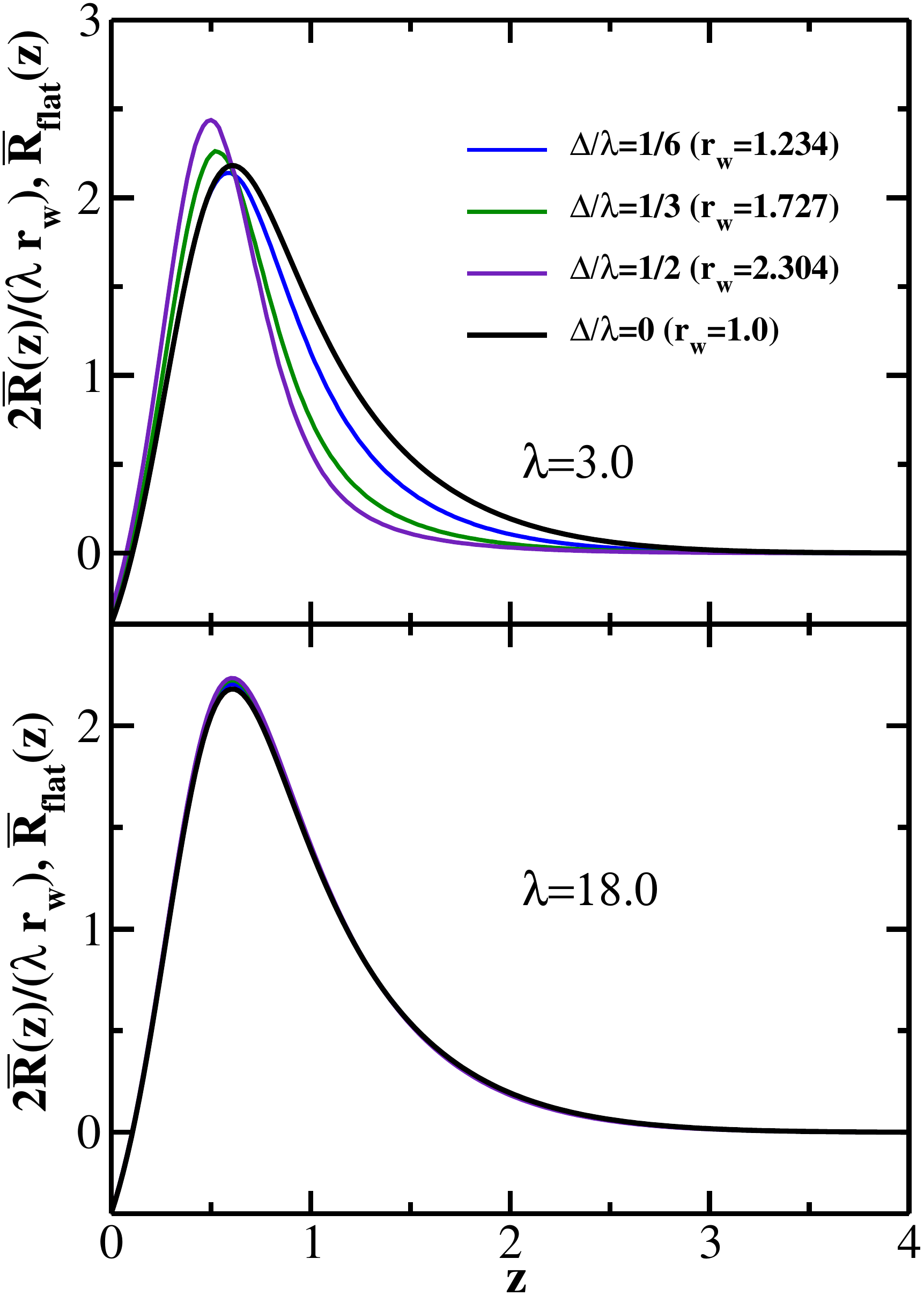} 
\caption{Function $2\bar{R}(z)/(\lambda r_{\rm w})$ (defined in the text) vs $z$ for several values of the Wenzel's ratio $r_{\rm w}$ as indicated. Upper panel: substrate wavelength $\lambda=3$, lower panel: substrate wavelength $\lambda=18$; black line indicates the result for a flat substrate.} 
\label{fig10}
\end{figure}

\begin{figure} 
\includegraphics[scale=0.5]{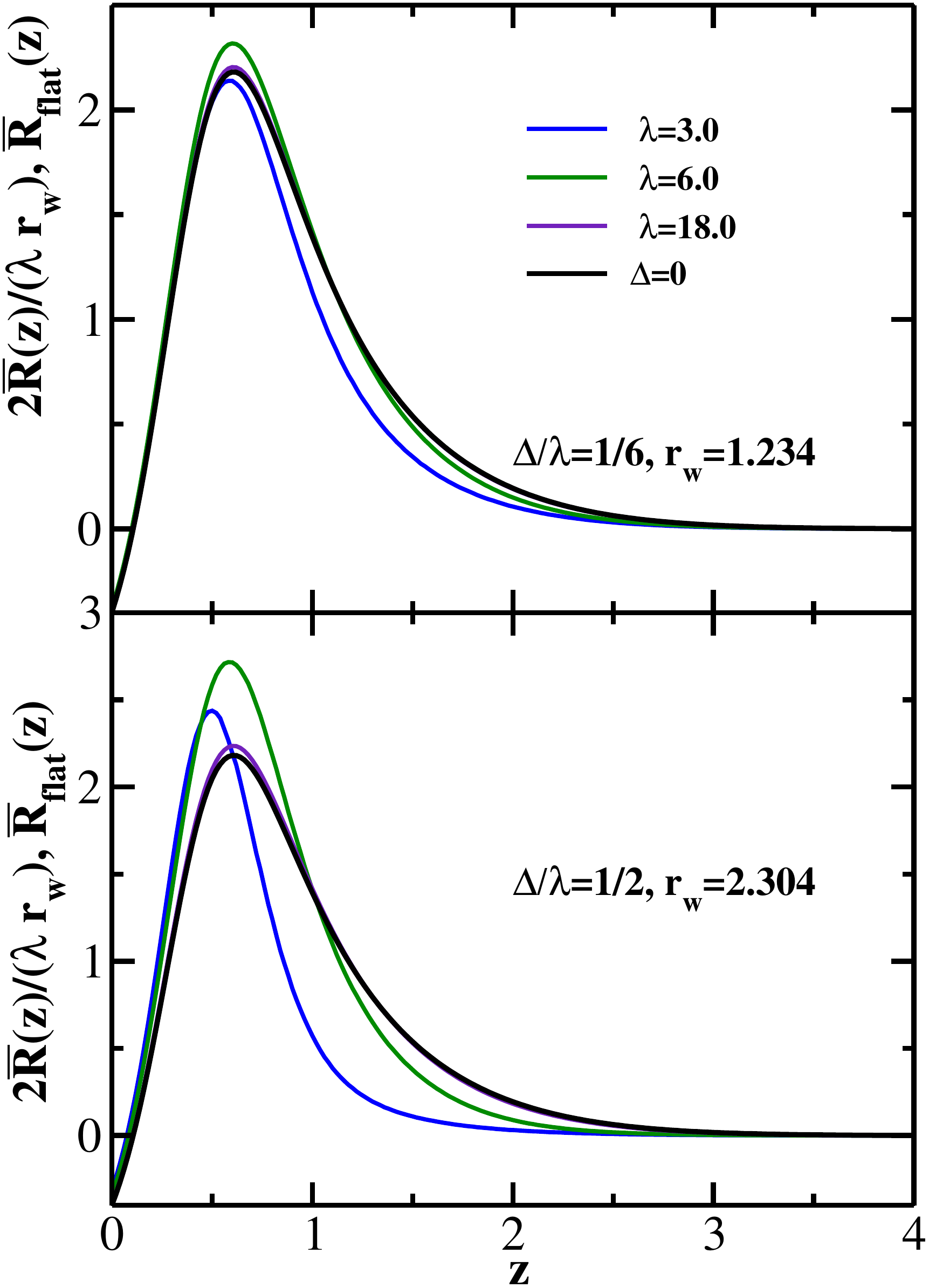} 
\caption{Function $2\bar{R}(z)/(\lambda r_{\rm w})$ (defined in the text) vs $z$ for several values of the 
substrate wavelength as indicated. Upper panel: substrate wavelength $r_{\rm w}=1.234$, lower panel:  $r_{\rm w}=2.304$; black line indicates the result for a flat substrate.} 
\label{fig11} 
\end{figure}

In Fig.~\ref{fig10} we plot our numerical results for $2\bar{R}(z)/(\lambda r_{\rm w})$ vs $z$ for several values of the
ratio $\Delta/\lambda$ (each of which corresponds to a particular value of $r_{\rm w}$:  
$\Delta/\lambda=1/6$ ($r_{\rm w}=1.234$), $\Delta/\lambda=1/3$ ($r_{\rm w}=1.727$),
and $\Delta/\lambda=1/2$ ($r_{\rm w}=2.304$); also included is the result for a flat profile
($\Delta=0$, $r_{\rm w}=1$). Upper panel presents the results for $\lambda=3$, and lower panel -- for $\lambda=18$.
In Fig.~\ref{fig11} we plot our numerical results for $2\bar{R}(z)/(\lambda r_{\rm w})$ vs $z$ for several values of the 
substrate wavelength: $\lambda=3$, $\lambda=6$, and $\lambda=18$. Upper panel presents the results for
$r_{\rm w}=1.234$ ($\Delta=\lambda/6$), and lower panel -- for $r_{\rm w}=2.304$ ($\Delta=\lambda/2$); also included in both panels is the result for a flat profile ($r_{\rm w}=1$). From these two figures one sees that the deviation of 
$2\bar{R}(z)/(\lambda r_{\rm w})$ from $\bar{R}_{\rm flat}(z)$ decreases with decreasing corrugation and increasing substrate wavelength, as one would expect. Overall, the deviations from Wenzel's relation arise from two sources: first, the dependence of the integrated gas-substrate potential on the parameter $t$ along the curves $z_g(t)$ and  $z_o(t)$ (which is implicitly ignored in Wenzel's relation), and second, the difference between the arc-lengths of $z_g(t)$ and  $z_o(t)$ (also ignored in Wenzel's relation), as already discussed in the previous Section, by comparing Eqs.~(\ref{ratioroughflat}) and (\ref{ratioroughflatwenzel}). Both these deviations eventually disappear in the macroscopic limit.


\begin{figure} 
\includegraphics[scale=0.5]{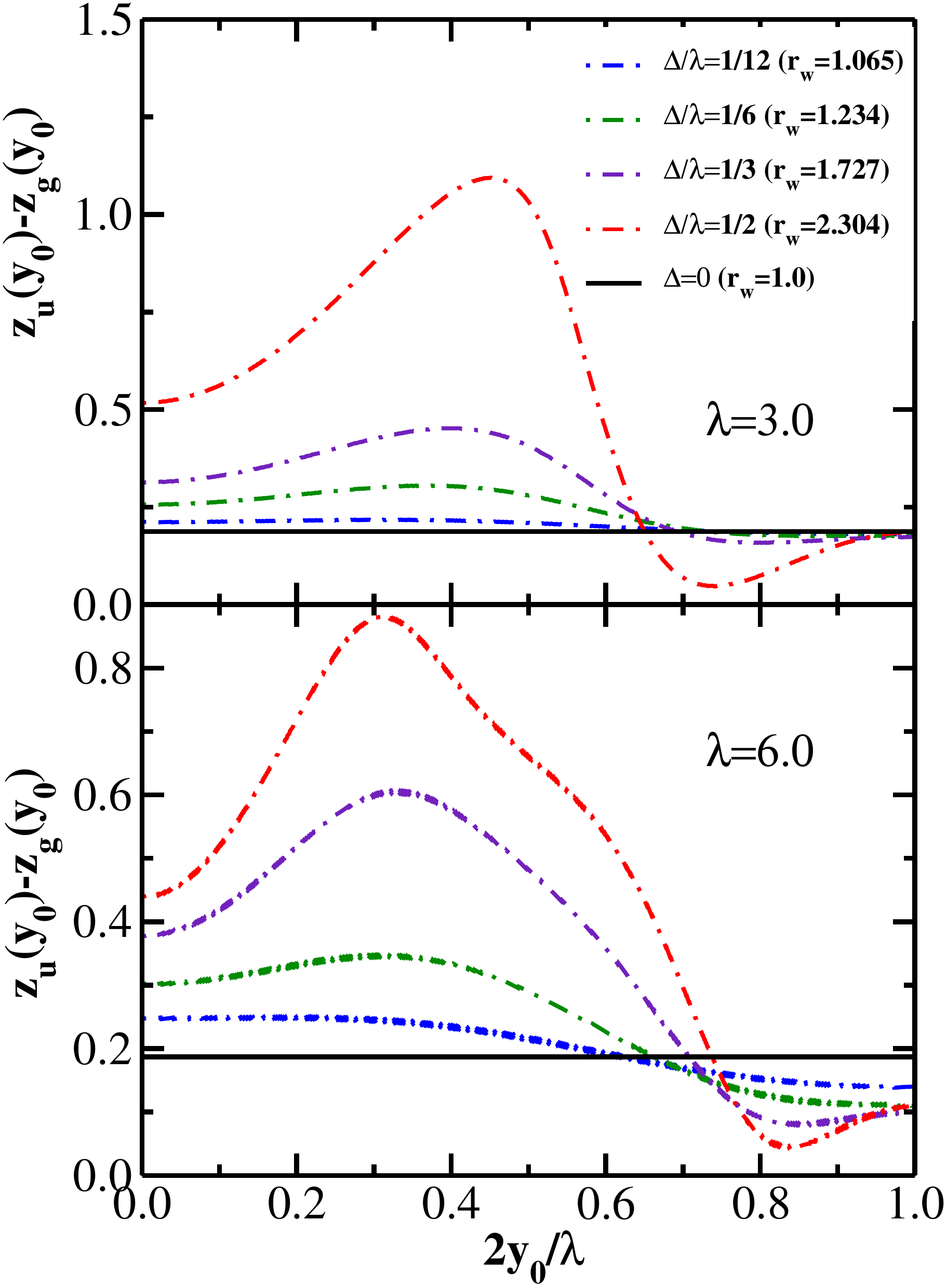} 
\caption{Function $z_u(y_0)-z_g(y_0)$ (where $z_u(y_0)$ is the value of $z_0$ at which the external potential has the chosen value $\beta u_{\rm ext}(y_0,z_0)=-0.35$) vs $2y_0/\lambda$ for several values of the 
Wenzel's ratio as indicated. Upper panel: substrate wavelength $\lambda=3.0$, lower panel:  $\lambda=6.0$; black line indicates the result for a flat substrate: $z_{u}^{\rm flat}=0.187$, see Fig.~\ref{fig7}.} 
\label{fig12} 
\end{figure}

\begin{figure} 
\includegraphics[scale=0.5]{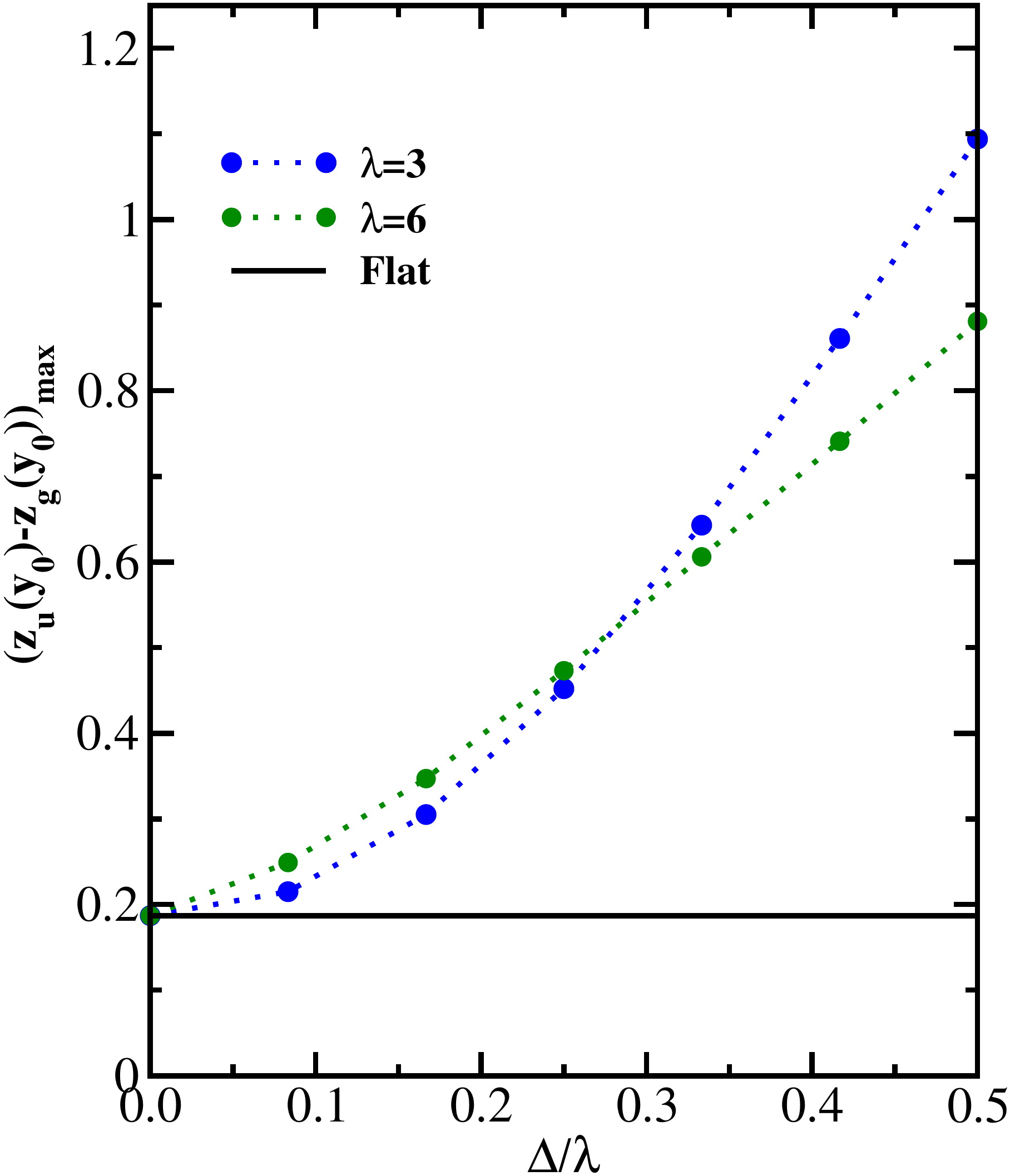} 
\caption{Maximum value of the function $z_u(y_0)-z_g(y_0)$ from Fig.~\ref{fig12} vs the ratio $\Delta/\lambda$ for two values of the substrate wavelength, as indicated; black line indicates the result for a flat substrate.} 
\label{fig13} 
\end{figure}

In order to illustrate the first source of deviations in greater detail, we have chosen a specific value of the external potential, $\beta u_{\rm ext}(y_0,z_0)=-0.35$, and have computed the corresponding value of the coordinate $z_0=z_u(y_0)$ where the external potential has the chosen value as a function of $y_0$ from  $y_0=0$ to
$y_0=\lambda/2$. Note that there are two values of $z_u(y_0)$ corresponding to the above condition (before and after the minimum of the external potential); we have chosen the smaller value, i.e. the one closer to the substrate, and present it as the difference from the height of the substrate at the same value of $y_0$, $z_u(y_0)-z_g(y_0)$. Our results for $z_u(y_0)-z_g(y_0)$ as a function of $2y_0/\lambda$ are shown in Fig.~\ref{fig12} for two values of the substrate wavelength: $\lambda=3$ (upper panel) and      
$\lambda=6$ (lower panel). As in the previous figures, the results are presented for several values of the
ratio $\Delta/\lambda$, each corresponding to a particular value of $r_{\rm w}$, as indicated. One sees that for the largest amplitude-to-wavelength ratio considered here ($\Delta/\lambda=1/2$) the deviation of  $z_u(y_0)-z_g(y_0)$ from its value for a flat substrate is rather strong, and it gradually decreases with decreasing ratio
$\Delta/\lambda$. This gradual convergence to the Wenzel limit is shown in Fig.~\ref{fig13}, where we plot the maximum value of the function $z_u(y_0)-z_g(y_0)$ from Fig.~\ref{fig12} vs the ratio $\Delta/\lambda$ for two values of the substrate wavelength: $\lambda=3$ and $\lambda=6$; also shown is the corresponding result for the flat substrate, $\Delta=0$. While in the limit $\Delta/\lambda\rightarrow 0$ the Wenzel regime (which ignores the deviation from the flat substrate result) is indeed approached, this approach is once again non-monotonic:
for $\Delta/\lambda<1/4$, the deviation for $\lambda=6$ is larger compared to $\lambda=3$. 

\begin{figure} 
\includegraphics[scale=0.5]{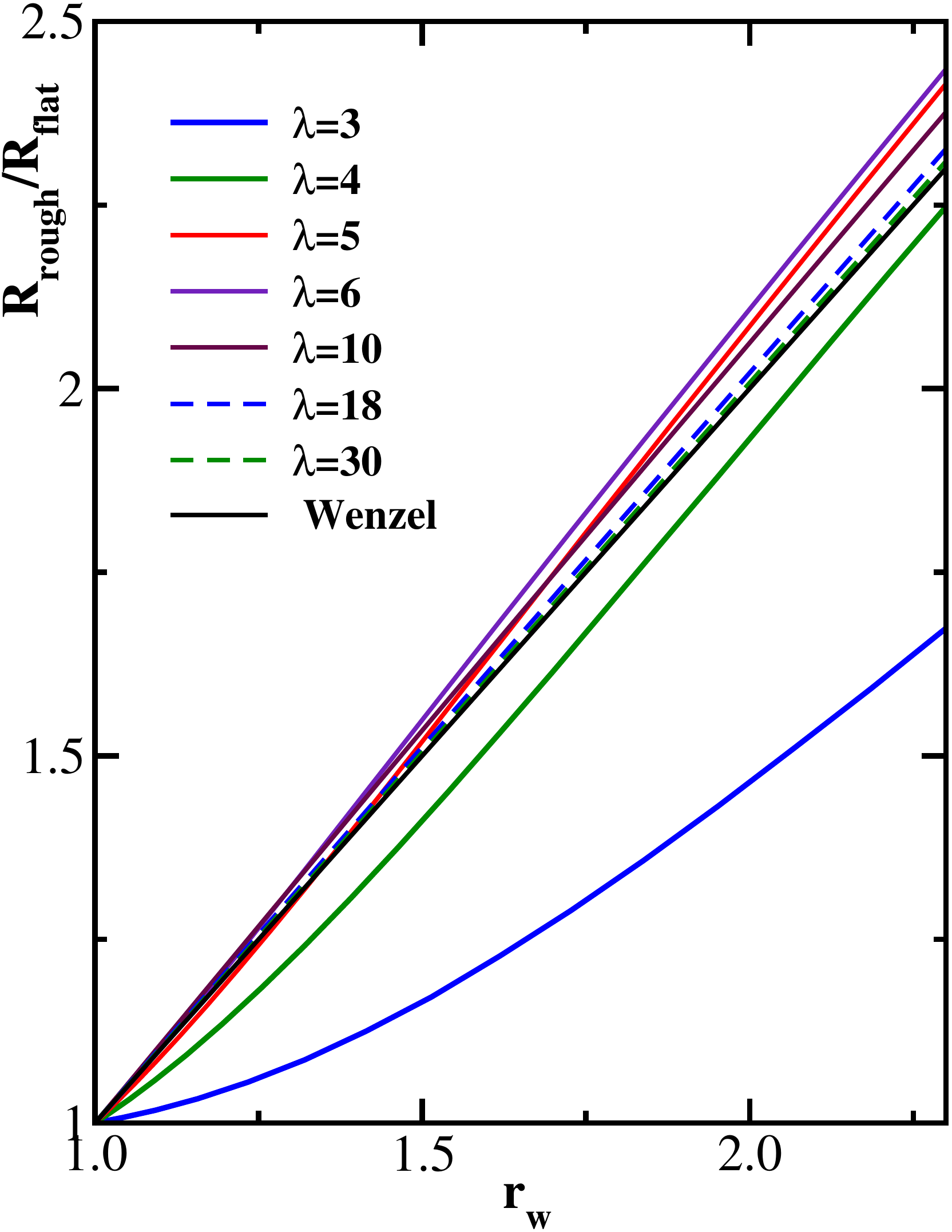} 
\caption{The ratio $R_{\rm rough}/R_{\rm flat}$ from Eq.~(\ref{gammasvratiogauss}) vs the Wenzel's ratio $r_{\rm w}$ for several values of $\lambda$, as indicated. The black line indicates the Wenzel prediction, which is straight line with slope 1. Recall that $\lambda$ is 
measured in units of $d_x=1$, the parameter characterizing the range of
the wall potential (see Eq.~(\ref{uextyz}) and Fig.~\ref{fig7}).} 
\label{fig14} 
\end{figure}

Finally, to summarize the deviations from Wenzel's relation considered in this Section in one
graph (similar to Fig.~\ref{fig4} in Sec.~\ref{subsection2.1}), we compute
the ratio of the gas-wall interfacial tensions for the rough and flat surfaces:~\cite{sitta16}
\begin{equation}
  \frac{\gamma_{sv}^{\rm rough}}{\gamma_{sv}}=\frac{\int_{0}^{\infty} dz 2R(z)/\lambda}{\int_{0}^{\infty} dz
    R_{\rm flat}(z)}=\frac{R_{\rm rough}}{R_{\rm flat}},
  \label{gammasvratiogauss}
\end{equation}
where the values of $R_{\rm rough}=\int_{0}^{\infty} dz 2R(z)/\lambda$ are given by the areas under the corresponding curves in Figs.~\ref{fig10} and~\ref{fig11} multiplied by $r_{\rm w}$. Our numerical results for the ratio
$R_{\rm rough}/R_{\rm flat}$ are shown as a function of $r_w$ in Fig.~\ref{fig14} (Wenzel's relation simply predicts a straight line with the slope of unity). The results are given for several values of the substrate wavelength and the Wenzel behavior is gradually approached with increasing $\lambda$: the results for $\lambda=30$ are nearly in the Wenzel limit for all values of $r_{\rm w}$ considered here. However, as one could already expect from Figs.~\ref{fig9} and~\ref{fig13}, this approach occurs non-monotonically, with the results for $\lambda=3$ and $\lambda=4$ lying below the Wenzel line, while the results for larger wavelengths are all above the Wenzel line. Given that the approach to the Wenzel limit in the case of the two-Gaussian potential illustrated in Fig.~\ref{fig14} differs significantly from the case of the square-shoulder potential shown in Fig.~\ref{fig4}, one would not expect the scaling relation depicted in Fig.~\ref{fig6} to hold also in the present case. Indeed, we have re-plotted our data from Fig.~\ref{fig14} in the form of Fig.~\ref{fig6}, and did not observe the above scaling to hold (not shown).

In summary, we have considered ideal gas in contact with a corrugated substrate (with two different models for the gas-substrate interaction) and using this simple model illustrated the origin of deviations from the Wenzel relation.
The behavior shown in Figs.~\ref{fig3}-\ref{fig14} is unexpectedly rich and many interesting details can be pointed out; for instance, although density profiles along offset curves can differ singnificantly from the values for the planar case, the average can be close, particularly for inmtermediate values of $\lambda$. However, more studies for different potentials would be required to clarify whether such details are general features or specific for the potentials chosen in the present study.
In the next Section, we will go beyond the ideal gas and use DFT to compute the gas-solid and liquid-solid surface tensions of an interacting fluid at a corrugated substrate, compute the corresponding contact angle, and demonstrate the deviations from Wenzel's relation for this more realistic model.

\section{Density Functional Theory}
\label{section3}

As our microscopic model, we consider a non-additive binary mixture of soft penetrable spheres~\cite{kim12} where the particles of the two species ($i,j=1,2$) interact via bounded spherically symmetric potential defined by: 
\begin{equation} 
\phi_{ij}(r) = \epsilon_{ij}
\quad {\rm for} \quad 0\le r \le \sigma_{ij},
\quad\phi_{ij}(r) = 0
\quad {\rm for} \quad r> \sigma_{ij},
\label{phiij}
\end{equation}
where $\sigma_{ij}$ is the size parameter, and $\epsilon_{ij}$ is the strength of the bounded potential when the two spheres overlap. In what follows, we set $\epsilon_{11}=\epsilon_{22}=\epsilon_{12}=\epsilon_{21}=\epsilon=1$ as our energy unit, and $\sigma_{11}=\sigma_{22}=\sigma=1$ as our length unit. The degree of non-additivity is governed by parameter $\delta$ defined by:
$\sigma_{12}=\sigma_{21}=0.5(\sigma_{11}+\sigma_{22})(1+\delta)$.
Taking $x_i$ to be the mole fraction of species $i$, the total number density of the mixture is given by
$\rho_t=\rho_1+\rho_2$, where $\rho_i=x_i\rho_t$ are densities of the two species. 
While our main
motivation for the choice of this (somewhat unconventional) model is
computational efficiency, we re-emphasize here that it may also be relevant
for certain colloidal systems, as already pointed out in the Introduction~\cite{lekkerkerker11}.

\begin{figure} 
\includegraphics[scale=0.5]{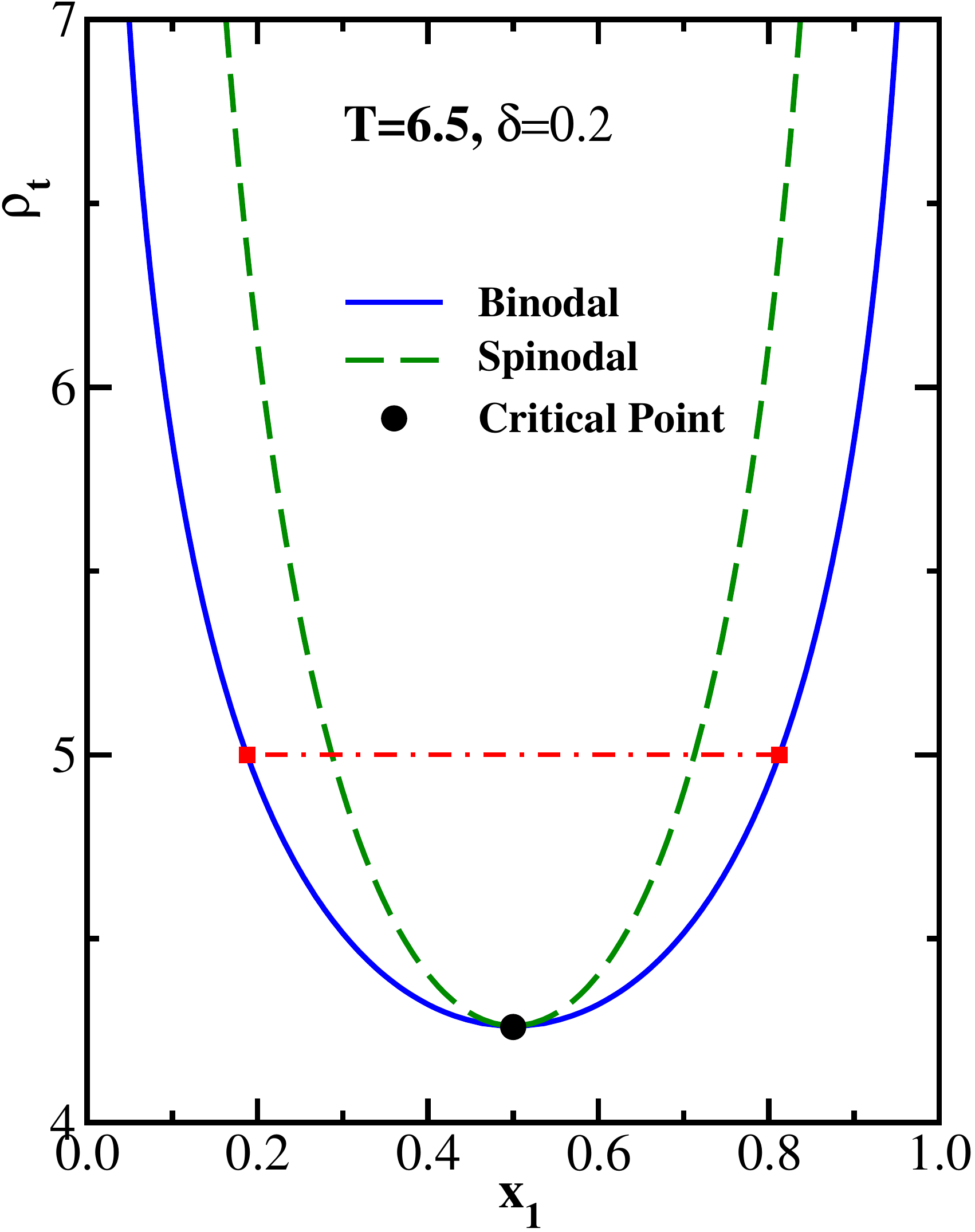} 
\caption{Phase diagram for the non-additive binary mixture in the variables total density $\rho_t$ vs the mole fraction of the first component $x_1$ at the dimensionless temperature $k_BT/\epsilon=6.5$ and non-additivity parameter $\delta=0.2$. Solid blue line is the binodal, dashed green line is the spinodal, critical point is indicated by a black dot, and the red dot-dashed tie-line connects the coexisting equilibrium densities at $\rho_t=5.0$ (marked by symbols) at which the interfacial tensions will be studied in this work: $\rho_{1}^{v}=\rho_{2}^{l}=0.94$ and $\rho_{1}^{l}=\rho_{2}^{v}=4.06$} 
\label{fig15} 
\end{figure}

Within the framework of the DFT, the (dimensionless) Helmholtz free energy of the mixture $\beta F$ 
is written as a sum of ideal and excess terms, where the former is known exactly:
\begin{equation}
\beta F_{\rm ideal}=\sum_{i=1}^{2}\int d\vec{r}\rho_i(\vec{r})[\ln(\Lambda_{i}^3\rho_i(\vec{r}))-1],
\label{fideal}
\end{equation}
while the latter is obtained in the mean-field approximation:
\begin{equation}
  \beta F_{\rm excess}=\frac{\beta}{2}\sum_{i,j=1}^{2}\int d\vec{r}\rho_i(\vec{r})\int d\vec{r}^{\prime}\rho_j(\vec{r}^{\prime}) \phi_{ij}(|\vec{r}-\vec{r}^{\prime}|),
        \label{fexcess}
\end{equation}
where $\rho_i(\vec{r})$ is the (inhomogeneous) density profile of species $i$, and $\Lambda_{i}$ is its 
de Broglie thermal length.  
Before considering the inhomogeneous case, it is important to establish the bulk phase diagram of the binary mixture, whereby the density profiles in the Eqs.~(\ref{fideal}) and (\ref{fexcess}) are taken to be uniform in space. The phase diagram is constructed using the standard procedure~\cite{kim12}, where the binodal curve (equilibrium densities at coexistence) is obtained by imposing the equality of pressure and chemical potentials of both components in the two coexisting phases, while the spinodal curve (limit of stability) is obtained from the divergence of compressibility of the mixture. The corresponding results obtained at the dimensionless temperature $k_BT/\epsilon=6.5$ and non-additivity parameter $\delta=0.2$ are shown in Fig.~\ref{fig15} in the variables total density $\rho_t$ vs the mole fraction of the first component $x_1$. In order to cast the discussion in the language of vapor-liquid coexistence, in what follows we will (arbitrarily) designate the phase rich in component 1 as ``liquid'' and the phase poor in component 1 as ``vapor'' for this symmetric binary mixture. In  Fig.~\ref{fig15} binodal is shown as a solid line, spinodal as a dashed line, and the location of the critical point ($x_{1}^{c}=0.5$, $\rho_{t}^{c}=4.263$) is marked by a circle. For the studies of interfacial tensions detailed below, we choose the total density sufficiently above the critical point ($\rho_t=5.0$ shown as dot-dashed tie-line in Fig.~\ref{fig15})) and the corresponding coexisting equilibrium densities are marked as symbols:
$\rho_{1}^{v}=\rho_{2}^{l}=0.94$ and $\rho_{1}^{l}=\rho_{2}^{v}=4.06$.  

Having established the bulk thermodynamics of the non-additive mixture, we now proceed to consider inhomogeneous systems. To this end, one defines the grand potential:
\begin{equation}
  \beta\Omega=\sum_{i=1}^{2}\int d\vec{r}\rho_i(\vec{r})[\ln(\Lambda_{i}^3\rho_i(\vec{r}))-1]+
  \frac{\beta}{2}\sum_{i,j=1}^{2}\int d\vec{r}\rho_i(\vec{r})\int d\vec{r}^{\prime}\rho_j(\vec{r}^{\prime}) \phi_{ij}(|\vec{r}-\vec{r}^{\prime}|)
  +\beta\sum_{i=1}^{2}\int d\vec{r}\rho_i(\vec{r})[\phi_{i}^{\rm ext}(\vec{r})-\mu_i],
  \label{betaomega}
\end{equation}
where $\phi_{i}^{\rm ext}(\vec{r})$ is the external potential acting on species $i$ and $\mu_i$ is its chemical potential. By minimizing $\beta\Omega$ one obtains the equilibrium density profiles in inhomogeneous systems. 
In particular, the inhomogeneous DFT calculations reported below are carried out on a Cartesian grid with the spacing $dz=0.02$ (in 1-d case) and $dy=dz=0.02$ (in 2-d case), with numerical integration performed using 2-point Gaussian quadrature and employing simple Picard iterative procedure, which was found to be adequate for the present simple microscopic model.

\begin{figure} 
\includegraphics[scale=0.5]{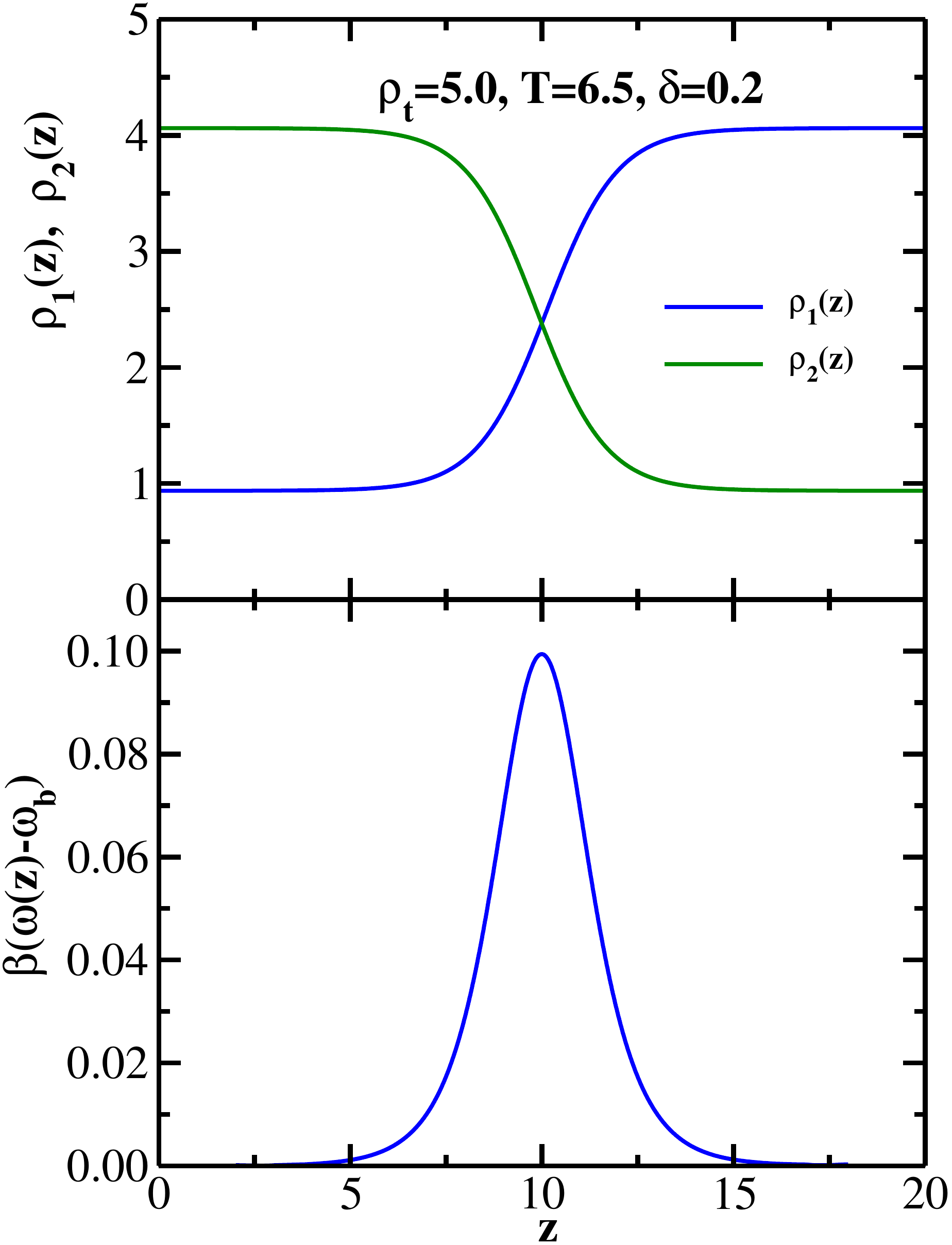} 
\caption{Upper panel: equilibrium density profiles at liquid-vapor coexistence at the total density $\rho_t=5.0$, dimensionless temperature $k_BT/\epsilon=6.5$ and non-additivity parameter $\delta=0.2$.
  Lower panel: dimensionless grand potential density relative to its bulk value as a function of $z$ across the liquid-vapor interface.} 
\label{fig16} 
\end{figure}

We start by considering a planar liquid-vapor interface located in $xy$-plane, in which case the density profiles depend on the $z$-coordinate only. We set $\rho_t=5.0$, $k_BT/\epsilon=6.5$, $\delta=0.2$, and compute the equilibrium density profiles at coexistence ($\rho_{1}^{v}=\rho_{2}^{l}=0.94$ and $\rho_{1}^{l}=\rho_{2}^{v}=4.06$), the DFT results are shown in the upper panel of Fig.~\ref{fig16}. Note that in this calculation
$\phi_{i}^{\rm ext}(\vec{r})$ and the boundary conditions are set such that one has bulk vapor phase at small $z$ and bulk liquid phase at large $z$. 
From these density profiles and Eq.~(\ref{betaomega}), one can readily compute the grand potential density $\beta\omega(z)$, which yields the liquid-vapor interfacial tension:
\begin{equation}
\beta\sigma^2\gamma_{lv}=\int_{-\infty}^{\infty}dz\beta(\omega(z)-\omega_b),
  \label{gammalv}
\end{equation}  
where $\beta\omega_b$ is the bulk value of the grand potential density. The integrand of Eq.~(\ref{gammalv}) is plotted in the lower panel of Fig.~\ref{fig16}, and the integration gives 
$\beta\sigma^2\gamma_{lv}=0.327$ at the state point considered. This value will be used in calculating the contact angle from Eq.~(\ref{young}).

\begin{figure} 
\includegraphics[scale=0.5]{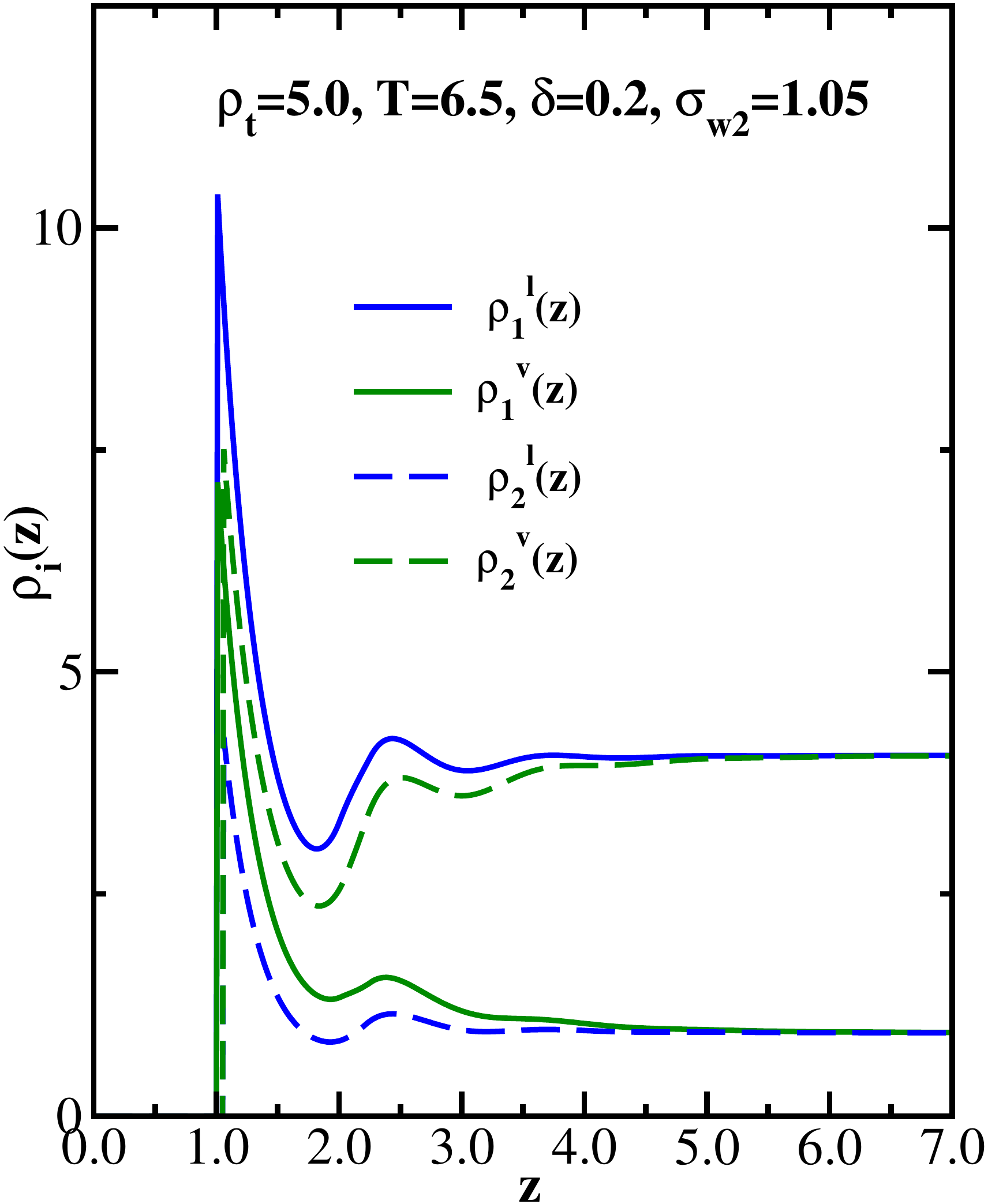} 
\caption{Equilibrium density profiles as a function of distance $z$ from the substrate at coexistence at the total density $\rho_t=5.0$, dimensionless temperature $k_BT/\epsilon=6.5$ and non-additivity parameter $\delta=0.2$; the external potential parameters are $\sigma_{\rm w1}=1.0$ and $\sigma_{\rm w2}=1.05$.
Solid lines are for component 1 and dashed lines are for component 2. Blue lines show liquid density profiles and green lines show vapor density profiles.} 
\label{fig17} 
\end{figure}

Proceeding next to the calculation of the solid-vapor and solid-liquid surface tensions, we start with the case of a flat substrate (located in $xy$-plane) and define the external potential similar to the one we used in Section~\ref{subsection2.1}:
\begin{equation} 
\phi_{i}^{\rm ext}(z) = \infty
\quad {\rm for} \quad z < \sigma_{\rm wi},
\quad \phi_{i}^{\rm ext}(z) = 0
\quad {\rm for} \quad z\ge \sigma_{\rm wi}.
\label{phiiext}
\end{equation} 
We note here that in our model the wetting properties of the (planar) wall are governed not by the temperature (which is held fixed in all the calculations at the value specified in the bulk phase diagram in Fig.~\ref{fig15}), but rather by the relative widths $\sigma_{wi}$ of the wall square shoulder potential for the two components of the binary mixture.
In what follows, we set the wall parameter for the first component $\sigma_{\rm w1}=1.0$ and control the contact angle by varying the wall parameter  $\sigma_{\rm w2}$ for the second component. Once again, we set $\rho_t=5.0$, $k_BT/\epsilon=6.5$, $\delta=0.2$, and compute the equilibrium density profiles as a function of distance $z$ from the substrate at coexistence ($\rho_{1}^{v}=\rho_{2}^{l}=0.94$ and $\rho_{1}^{l}=\rho_{2}^{v}=4.06$). Our DFT results for $\sigma_{\rm w2}=1.05$ are shown in Fig.~\ref{fig17}, with blue lines showing liquid density profiles and green lines showing vapor density profiles. Solid lines correspond to component 1 and dashed lines -- to component 2. One sees that the peaks in the density profiles of the first component are higher compared to the second component, as one would expect from the fact that $\sigma_{\rm w2}>\sigma_{\rm w1}$. Given that the liquid phase is enriched in component 1, one would expect that $\gamma_{sv}>\gamma_{lv}$ for the above values of parameters.
In order to confirm this, we compute $\gamma_{sv}$ ($\gamma_{lv}$) from the equilibrium density profiles by setting the boundary condition far away from the wall to vapor (liquid) phase and performing the integral 
$\int_{0}^{\infty}dz\beta(\omega(z)-\omega_b)$.
This procedure yields $\beta\sigma^2\gamma_{sv}=11.969$ and $\beta\sigma^2\gamma_{sl}=11.789$. 
Substituting these interfacial tensions and $\beta\sigma^2\gamma_{lv}=0.327$ into Eq.~(\ref{young}), we compute the contact angle and find $\cos\theta=0.55$. Thus, with the above choice of parameters we are sufficiently far removed from the wetting transition, and can use these parameters to study the effect of sinusoidal corrugation on the contact angle.  

\begin{figure}[htb]
\includegraphics[scale=0.3, angle=0]{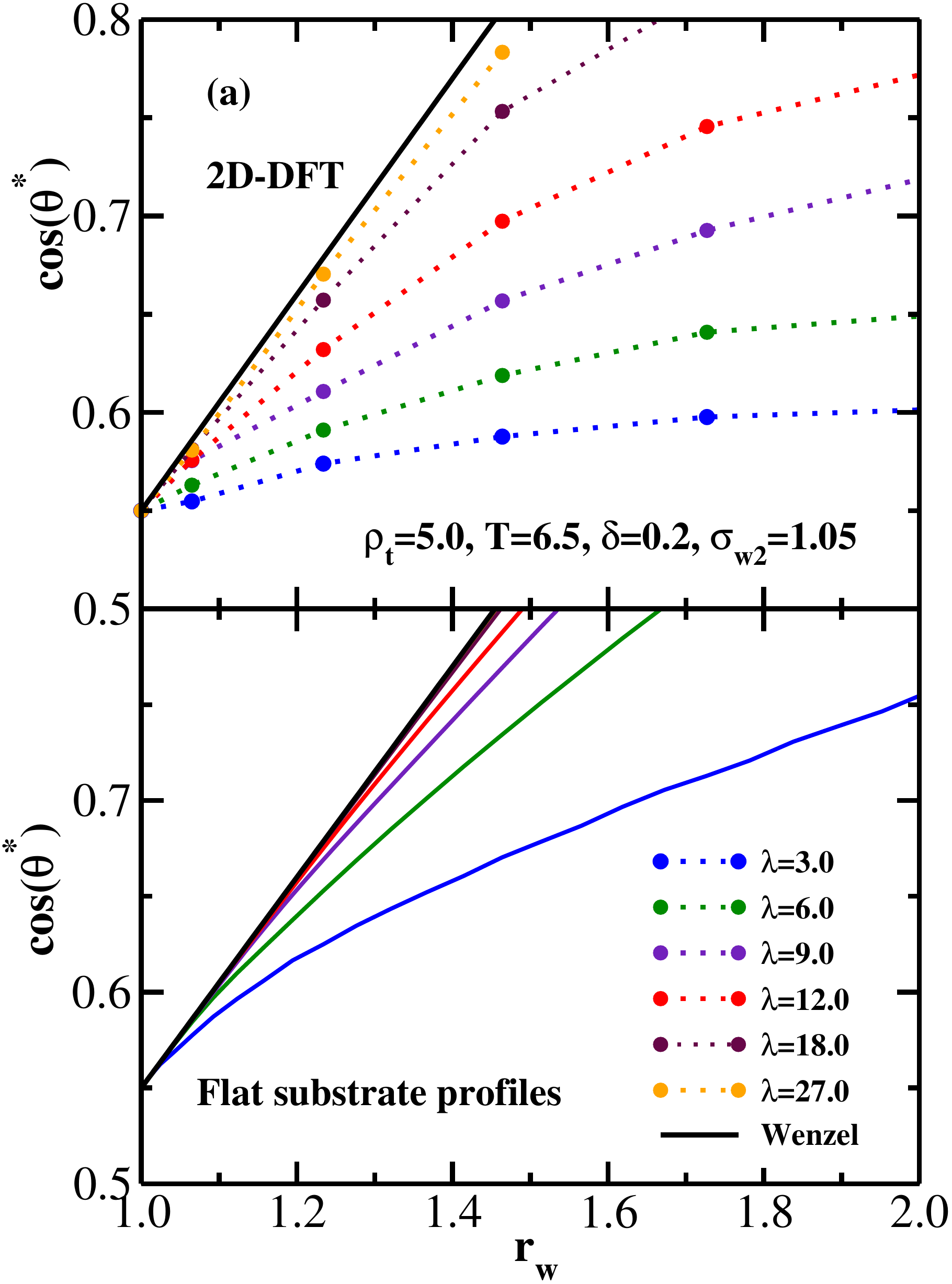}
\hspace{0.9cm}
\includegraphics[scale=0.3, angle=0]{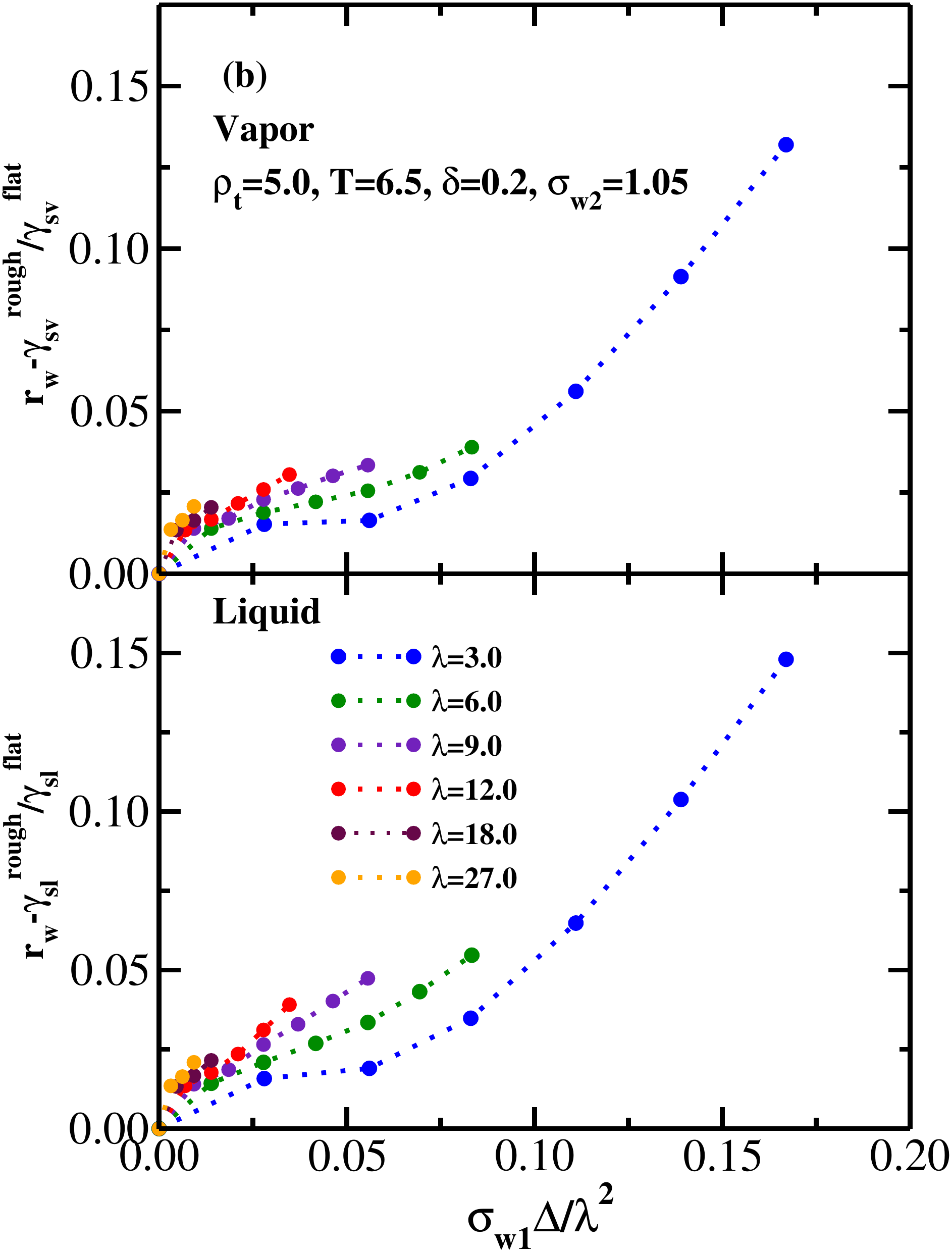}

\caption{(a) Upper panel: 2D-DFT results for the cosine of the contact angle $\theta^*$ as a function of the Wenzel parameter $r_{\rm w}$ for several values of the substrate wavelength $\lambda$, as indicated. Note that we only consider the regime $\cos(\theta^*)\le 0.8$ in order to avoid dealing with filling of the grooves of the substrate profile~\cite{rodriguezrivas15}.  
  Lower panel: same as the upper panel, except that 1D-DFT results for $\rho_i(z)$ were used in calculating the interfacial tensions, as explained in the text.
(b) The correction to the Wenzel's relation as a function 
 of dimensionless ratio $\sigma_{\rm w1}\Delta/\lambda^2$ for several values of $\lambda$, as indicated; the offsetFig.~\ref{fig18}a distance is fixed at $\sigma_{\rm w1}=1.0$. Upper panel: vapor (the correction plotted is $r_{\rm w}-\gamma_{\rm sv}^{\rm rough}/\gamma_{\rm sv}^{\rm flat}$); lower panel: liquid (the correction plotted is $r_{\rm w}-\gamma_{\rm sl}^{\rm rough}/\gamma_{\rm sl}^{\rm flat}$).} 
\label{fig18} 
\end{figure}

\begin{figure} 
\includegraphics[scale=0.5]{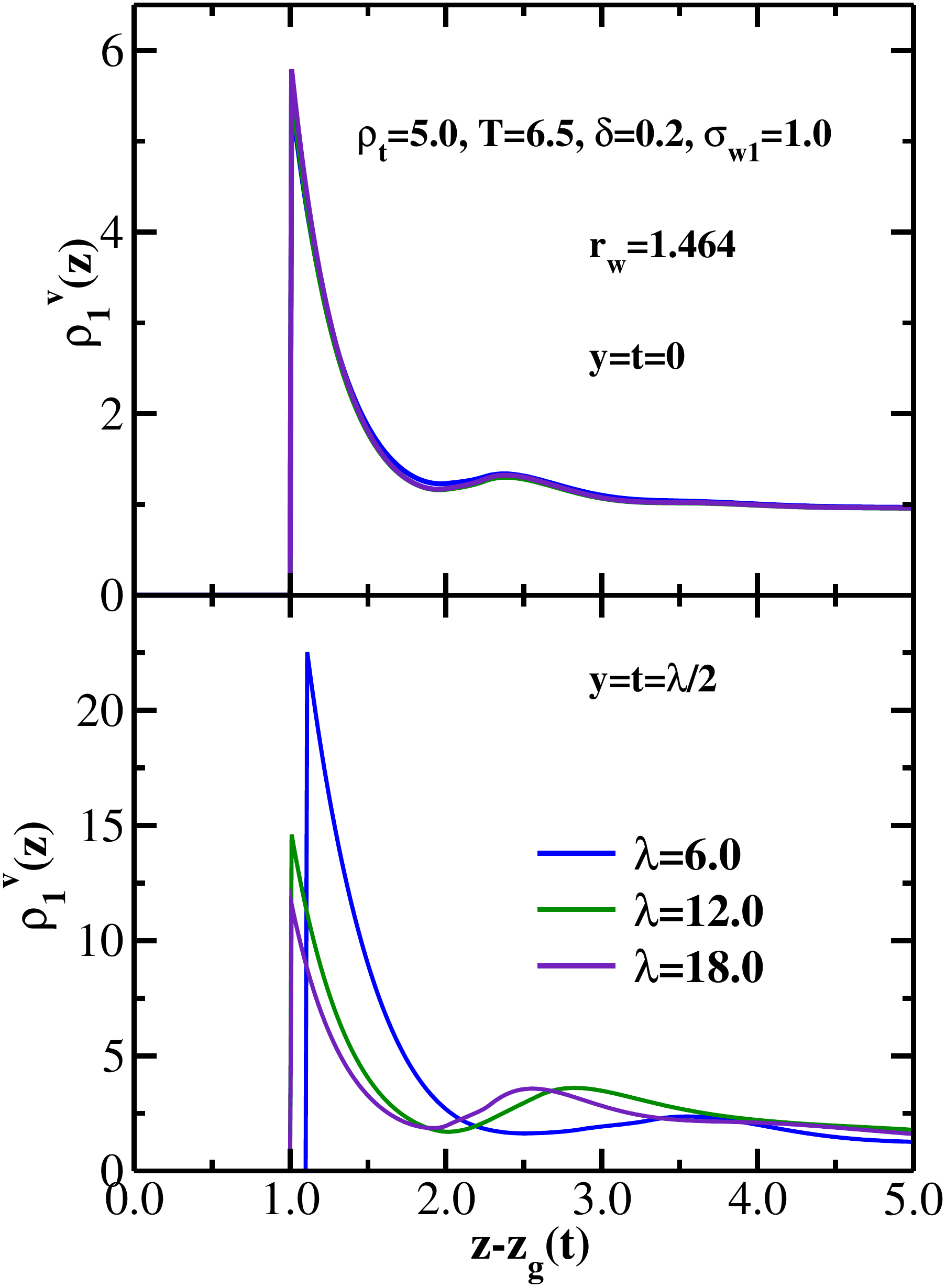} 
\caption{2D-DFT results for the density profiles of the first component in the vapor phase along the $z$-axis as a function of the distance from the substrate for three values of the wavelength as indicated. The Wenzel parameter value is $r_{\rm w}=1.464$ ($\Delta/\lambda=0.25$). Upper panel: $y=t=0$ (top of the cosine profile); lower panel:
$y=t=0.5\lambda$ (bottom of the cosine profile).} 
\label{fig19} 
\end{figure}

As in the previous section, we consider here  a weak sinusoidal corrugation in one direction only, i.e. the wall position $z_{\rm wall}$ relative to the planar reference wall at $z=0$ is given by
$z_{\rm wall}(y)=\Delta\sin(2\pi y/\lambda)$. The external potentials now are defined such that
$\phi_{i}^{\rm ext}(y,z)=\infty$ for all $(y,z)$ points whose closest distance from $z_{\rm wall}(y)$ is smaller than
$\sigma_{\rm wi}$ and $\phi_{i}^{\rm ext}(y,z)=0$ otherwise. All the parameters are taken to be the same as in the reference planar wall calculation above, and the equilibrium 2D-DFT density profiles $\rho_i(y,z)$ are obtained by minimizing the grand potential in 2 dimensions. From the resulting $\rho_i(y,z)$ one obtains the interfacial tensions and the contact angles as before. We have computed the contact angle for several values of the corrugation wavelength $\lambda$  as a function of the amplitude $\Delta$ and present our DFT results for $\cos\theta$ as a function of the corresponding Wenzel parameter $r_{\rm w}$. The results from 2D-DFT calculations are shown in the upper panel of Fig.~\ref{fig18} for six values of $\lambda$ together with the Wenzel's prediction. One sees that the latter consistently overestimates the contact angle, and the DFT results only start approaching the Wenzel limit for the largest wavelength considered here, $\lambda=27$. In order to illustrate the behavior of the equilibrium 2D-DFT density profiles $\rho_i(y,z)$, in Fig.~\ref{fig19} we present as an example the corresponding results for the first component in the vapor phase for several values of the wavelength $\lambda$. The profiles are shown as a function of the distance from the substrate along the $z$-axis for two particular values of $y$: $y=t=0$ (top of the cosine curve, upper panel) and $y=t=0.5\lambda$ (bottom of the cosine curve, lower panel). One observes that at the top of the substrate
the density profiles for all three wavelengths are quite similar to each other (and to the corresponding profile for the flat substrate shown in Fig.~\ref{fig17}), while at the bottom of the substrate there is a strong accumulation of the first component near the substrate (compared to the flat case), especially for the smallest wavelength $\lambda=6$. Although at the bottom of the groove right at the substrate a larger 
maximal density is reached than for the dense liquid phase in the case 
of a planar substrate (see Fig.~\ref{fig17}), the density profile for the shallow 
groove does not resemble that of a thin liquid film of a planar 
substrate: from Fig.~\ref{fig17} we would expect that then the density is about 
4.06 for $z > 3$ and stays constant for a range of $z$ (for a thin liquid domain),
but this is not what one sees in Fig.~\ref{fig19}. Obviously, the shallow 
groove provided by the minimum of the sinusoidal corrugation does not have enough space to accommodate a precursor of the liquid domain that one finds in macroscopic grooves where near the wetting transition
partial filling occurs~\cite{rodriguezrivas15}. 

The 2D-DFT calculations are rather demanding computationally, and hence, only a few selected values of the amplitude $\Delta$ were considered for each $\lambda$, as indicated by symbols on the DFT lines in the upper panel of Fig.~\ref{fig18}a. Accordingly, it is of interest to ask if one could utilize the (much cheaper) 1D-DFT density profiles $\rho_i(z)$ at a flat wall to construct some approximation for the two dimensional density distribution $\rho_i(y,z)$.  One such possibility is to approximate the density distributions along the lines {\em normal to the sinusoidal substrate} by the corresponding ``flat profiles'' $\rho_i(z)$. Given that the latter are obtained on a grid with some small spacing (e.g. $dz=0.01\sigma$) this procedure yields the values of $\rho_i(y,z)$ (and hence, the grand potential $\beta\omega(y,z)$) along the set of parallel curves (as defined in Sec.~\ref{section2}) spaced by the increment $dz$. By integrating $\beta(\omega(y,z)-\omega_b)$ along these curves and then along $z$ one obtains the corresponding interfacial tensions and the contact angle. The corresponding results are shown in the lower panel of Fig.~\ref{fig18}a, and one sees that they are qualitatively similar to 2D-DFT results, but the deviations from Wenzel's result are significantly underestimated in this approach, and therefore it cannot be considered as a viable alternative to full-scale 2D-DFT calculations.

Given the similarity of the external potential given by Eq.~(\ref{phiiext}) to the one used in
  Section~\ref{subsection2.1} for the ideal gas case, it would be of interest to ask to what extent the scaling relation illustrated in Fig.~\ref{fig6} is obeyed by the penetrable fluid. To this end, in the upper panel of Fig.~\ref{fig18}b we have plotted the correction to the Wenzel's relation for the vapor surface tension,
$r_{\rm w}-\gamma_{\rm sv}^{\rm rough}/\gamma_{\rm sv}^{\rm flat}$, 
  as a function 
 of dimensionless ratio $\sigma_{\rm w1}\Delta/\lambda^2$ for several values of $\lambda$; analogous results for the liquid case are plotted in the lower panel of Fig.~\ref{fig18}b. One sees that the aforementioned scaling is approximately followed by the DFT data, albeit not as closely as in the case of ideal gas in Fig.~\ref{fig6}.

\section{Summary}
\label{section4}

In this work we have studied the limits of applicability of Wenzel's extension of Young's equation for the contact angle of droplets on microrough surfaces.
To this end, we first considered ideal gas in contact with a corrugated substrate using two different models for the gas-substrate interaction and with this simple model illustrated the various origins of deviations from the Wenzel relation for the gas-solid surface tension.
This approach has the merit that the geometric origin of deviations between the surface tensions of the flat and corrugated substrates can be rigorously understood: (i) curves at a normal distance $\sigma_{\rm w}$ from the sinusoidal corrugation (Fig.~\ref{fig3}) have a different character  (and length) than the corrugation profile itself; (ii) the potential acting on a gas atom at a distance $h$ above a maximum of the corrugation profile differs from the potential at the same distance above a minimum, if it results from the summation of  a distance-dependent pairwise interaction of the gas particles with particles forming the substrate, see Figs.~\ref{fig8}-\ref{fig13}.
For a simple short-range wall potential, we found that the approach towards Wenzel's law with increasing wavelength $\lambda$ is slow and nonmonotonous. 

Next, we applied 2D-DFT to compute the gas-solid and liquid-solid surface tensions of an interacting fluid at a corrugated substrate, computed the corresponding contact angles, and demonstrated the deviations from Wenzel's relation for this more realistic model.
In typical cases, corrugation does
cause pronounced changes of the contact angle (Fig.~\ref{fig18}), although in most
cases the change is not as large as predicted by Wenzel's equation, 
except when the corrugation wavelength $\lambda$ is very large. Nontrivial changes of the local
density profiles in $z$-direction are predicted to occur due to the corrugation (Fig.~\ref{fig19}). 
Finally, in the Appendix we discussed the interface Hamiltonian treatment of wetting on sinusoidally corrugated substrates. This latter approach models the substrate surface in the partial wetting state as being coated with a very thin precursor of a liquid wetting layer, of nanoscopic thickness $l(y)$, see Eq.~(\ref{lysin}) and Fig.~\ref{fig20}a. It is energetically favorable to have an amplitude $A < \lambda$ for the corrugation of this liquid -vapor interface bound to the substrate. With this theory we were able to obtain a good fit of the available simulation data (Fig.~\ref{fig21}). 

\section{Acknowledgments}
\label{section5}
SAE acknowledges financial support from Alexander von Humboldt foundation and thanks Prof. Andrey Milchev for helpful discussions. He also thanks Mr. Emmit Pert for help with creating figures.  

\section{Data availability}
\label{section6}

The data that support the findings of this study are available from the corresponding author upon reasonable request.
\appendix
\section{Interface Hamiltonian Treatment of Wetting on Sinusoidally Corrugated Surfaces}

In the Appendix we develop interface Hamiltonian treatment of wetting on sinusoidally corrugated substrates. 
Essentially this approach was already formulated by Parry et al.~\cite{swain98b,rascon00b}
in a discussion of wetting transitions on corrugated 
substrates. Note also that this approach can be criticized on various
grounds~\cite{rejmer07}. Here, however, we are not attempting to study 
phase transitions (first or second order wetting versus "filling" or
"unbending" or thin to thick film transitions), 
but only wish to clarify under which conditions (length scale of the
corrugation, etc) corrections to Wenzel's equation become negligible. 
An interface Hamiltonian theory of contact angles on heterogeneous surface has already been given by Swain and Lipowsky~\cite{swain98}, but their treatment referred to heterogeneities on scales much larger than ``mesoscopic'' lengths such as the distance $l_{\rm min}$ where the phenomenological interface potential has its minimum. Disregarding their extensions to account for line tension effects, gravity acting on the liquid etc, they derived the Wenzel rule for the average contact angle $\theta_{\rm av}$
\begin{equation}
  \cos\theta_{\rm av}=r_{\rm w}\cos\theta,
  \label{costhetaav}
\end{equation}
where $\theta$ is the contact angle on a perfectly planar but otherwise identical substrate surface, and
$r_{\rm w}$ 
is the ratio of the true surface area of the rough or corrugated surface to the planar surface area.

For simplicity, we consider here  a weak sinusoidal corrugation in one direction only, i.e. the wall position 
$z_{\rm wall}$ relative to the planar reference wall at $z=0$ is given by
\begin{equation}
  z_{\rm wall}(y)=\Delta\sin(2\pi y/\lambda),
  \label{zwall}
\end{equation}
so $\Delta$ describes the amplitude and $\lambda$ the wavelength of the corrugation, and no dependence on the $x$-coordinate is considered. (Note that we use sine function here, while we used cosine in the main text, but the two forms are completely equivalent, of course.) 
For this case, the Wenzel factor $r_{\rm w}$ is given by:
\begin{equation}
  r_{\rm w}=\frac{1}{\lambda}\int_{0}^{\lambda} dy \sqrt{1+
    \left(\frac{2\pi\Delta}{\lambda}\right)^2\cos^2\left(\frac{2\pi y}{\lambda}\right)}.
  \label{rwsinus}
\end{equation}
In the limit of small $\Delta/\lambda$ this reduces to
\begin{equation}
  r_{\rm w}\approx 1 +(\pi\Delta/\lambda)^2, \quad \Delta/\lambda\rightarrow 0
  \label{rwapprox}
\end{equation}
When we have an interface at position $z=l(y)$ in the absence of a potential acting on it, the free energy cost due to this interface is (per unit length in $x$-direction):
\begin{equation}
  F_{\rm int}=\gamma_{lv}\int_{0}^{y_{L}} dy \sqrt{1+(dl(y)/dy)^2}.
  \label{fint}
\end{equation}

The integral in Eq.~\ref{fint} is simply the length of the line when one cuts the interface with the $yz$-plane, and hence $\gamma_{lv}$ is the interface tension of a planar liquid-vapor interface.

Unlike Swain and Lipowsky~\cite{swain98}, we restrict attention to the case where $dl(y)/dy\ll 1$ everywhere. Then the interface free energy per unit area of the $xy$-plane becomes:
\begin{equation}
  F_{\rm int}/y_L=\gamma_{lv}\left[1+\frac{1}{y_L}\int_{0}^{y_{L}} dy \frac{1}{2}(dl(y)/dy)^2\right].
  \label{fintyl}
\end{equation}
Of course, for a free interface in the absence of a wall potential equilibrium is described by $dl/dy=0$ everywhere, there is no average enhancement of the surface area describing the interface as a quasi-two-dimensional object.

Wetting phenomena in this treatment are described~\cite{dietrich88} by exposing the interface to a wall potential $V(l,y)$. Thus one describes partial wetting by a liquid layer in between the wall and the interface. Therefore, the interface free energy between the vapor phase and the wall, per unit area of a flat planar interface, neglecting possible effects due to nonlocality of the wall potential~\cite{parry04}, is
\begin{equation}
  \gamma_{wv}=\gamma_{lv}+\frac{1}{y_L}\int_{0}^{y_{L}} dy[\gamma_{wl}(y)+V(l,y)+\frac{\gamma_{lv}}{2}(dl(y)/dy)^2].
  \label{gammawvyl}
\end{equation}
Eq.~(\ref{gammawvyl}) is appropriate for a  chemically heterogeneous flat planar surface - then both the wall-liquid surface tension $\gamma_{wl}(y)$ and the wall potential $V(l,y)$ depend on $y$, and minimization of
$\gamma_{wv}$ with respect to the function $l(y)$ is a nontrivial problem. In the case of a planar homogeneous surface, however, $\gamma_{wl}(y)=\gamma_{wl}$ and $V(l,y)=V(l)$ are independent of $y$, and minimizing $\gamma_{wv}$ with  with respect to $l$ yields the equilibrium distance $l_{\rm min}$ of the interface distance from the wall:
\begin{equation}
  \gamma_{wv}=\gamma_{lv}+\gamma_{wl}+V(l_{\rm min}), \quad \frac{dV(l)}{dl}|_{l=l_{\rm min}}=0.
  \label{gammawvplanar}
\end{equation}

In the following, a form of $V(l)$ appropriate for a short-range wall potential and first-order wetting/drying transition is assumed~\cite{dietrich88}:
\begin{equation}
\tilde{V}(l)\equiv V(l)/\gamma_{lv}=-\delta\epsilon a_0e^{-\kappa l} -b e^{-2\kappa l} +ce^{-3\kappa l}, 
  \label{vtildel}
\end{equation}  
where $\kappa^{-1}$ describes the range of the wall potential, $a_0$, $b$, and $c$ are dimensionless constants, and $\delta\epsilon$ is a constant parameter describing the distance from the mean-field stability limit (``spinodal'') of the partially wet phase. Note that from Eqs.~(\ref{young}), (\ref{gammawvplanar}), and (\ref{vtildel})  one immediately concludes that
\begin{equation}
  \cos\theta=1+\tilde{V}(l_{\rm min}),
  \label{costheta}
\end {equation}
and using the abbreviation 
\begin{equation}
  x=\exp(-\kappa l_{\rm min})
  \label{xkappalmin}
\end {equation}
one readily finds from Eq.~(\ref{vtildel})
\begin{equation}
  x=b/(3c)+\sqrt{(b/(3c))^2+\delta\epsilon a_{0}/(3c)}.
\label{xbc}
\end{equation}

The first-order wetting transition occurs when $\theta=0$ and hence $\tilde{V}(l_{\rm min})=0$, i.e.  
\begin{equation}
  x_t=b/(2c), \quad \delta\epsilon_t=-b^2/(4c a_0).
  \label{xt}
\end{equation}
It is then convenient to rewrite both $x$ and $\tilde{V}(l_{\rm min})$ in terms of
$\epsilon-\epsilon_t\equiv\delta\epsilon-\delta\epsilon_t$,
\begin{equation}
  x=\frac{2}{3}x_t[1+\frac{1}{2}\sqrt{1+\frac{3xa_0}{x_{t}^{2}c}(\epsilon-\epsilon_t)}],
  \label{xdeltaeps}
\end{equation}
and
\begin{equation}
  \tilde{V}(l_{\rm  min})=6cx\left(\frac{1}{27}x_{t}^{2}-\frac{1}{27}x_{t}^{2}
    \sqrt{1+\frac{3a_0}{c}\frac{\epsilon-\epsilon_t}{x_{t}^{2}}}-\frac{\epsilon-\epsilon_t}{9}\frac{a_0}{c}\right). 
  \label{vtildelmin}
\end{equation}  
The condition $\theta=\pi$ ($\cos\theta=-1$) then yields the drying transition, so the value of
$(\epsilon-\epsilon_t)a_0/c$ for which a given choice of the two parameters $x_t$ and $c$ yield 
$\tilde{V}(l_{\rm min})=-2$ controls the range from wetting to drying in this model Hamiltonian.

\begin{figure}[htb]
\includegraphics[scale=0.3, angle=0]{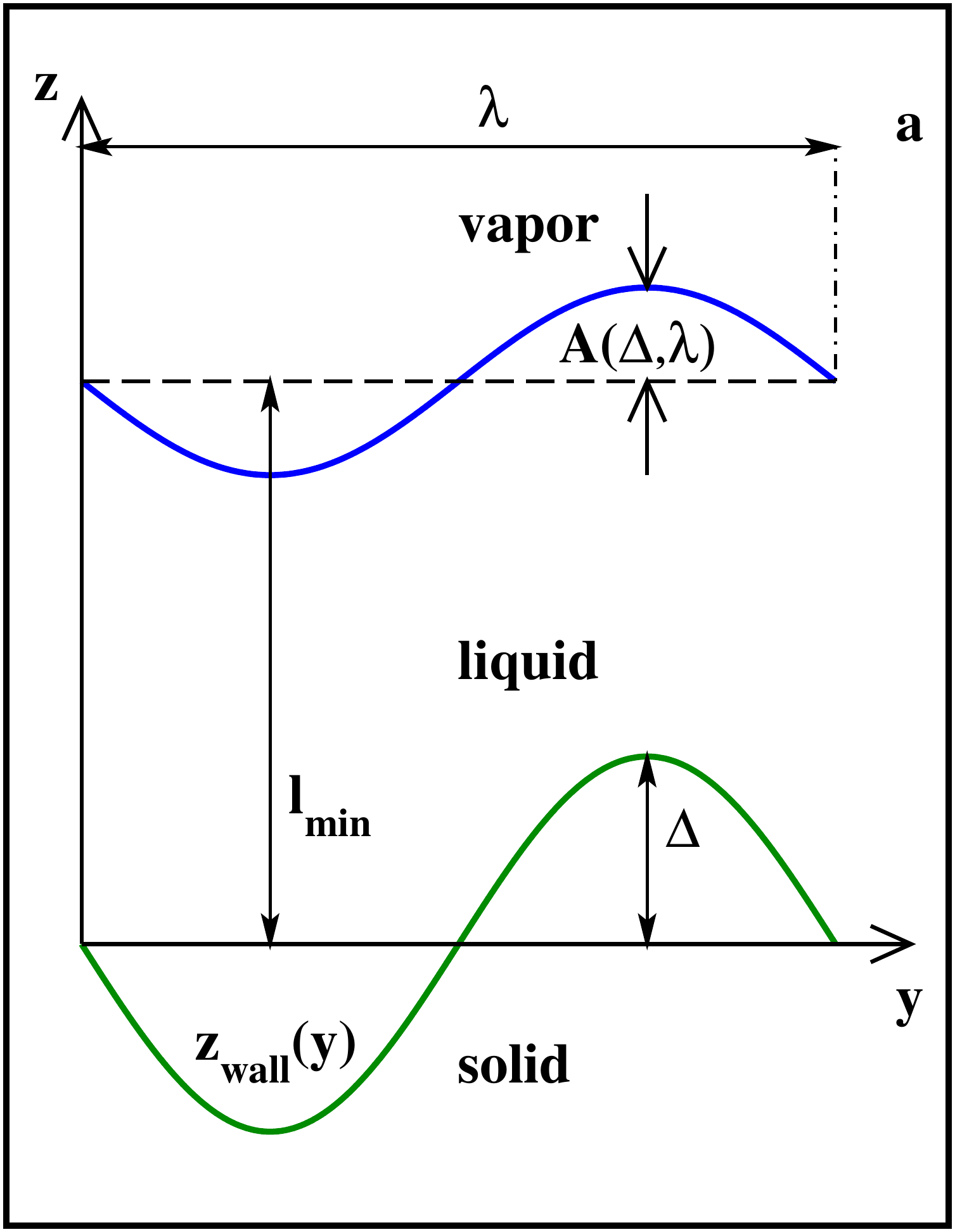} 
\hspace{0.9cm}
\includegraphics[scale=0.3, angle=0]{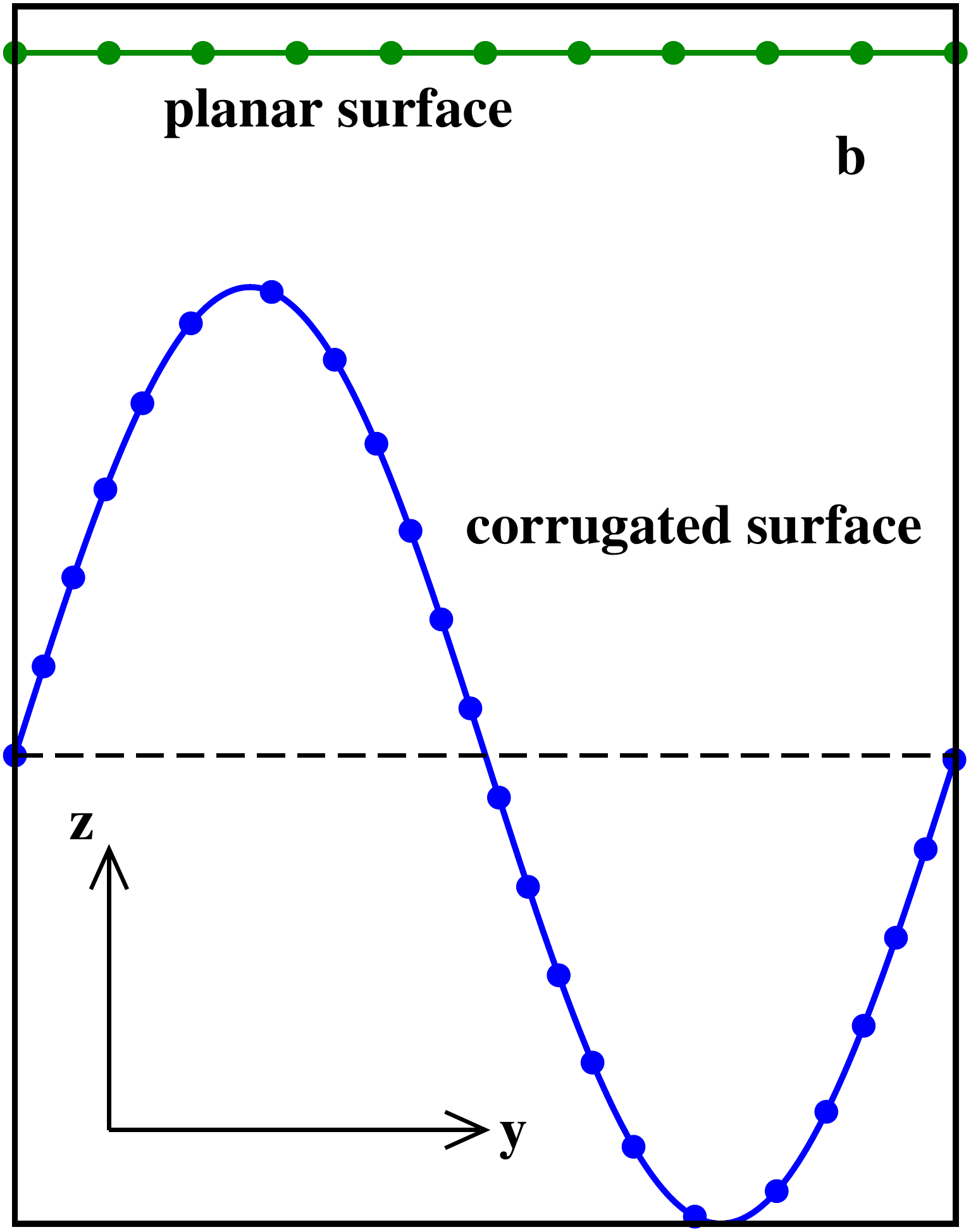}
\vspace{0.3cm}

\caption{(a) Schematic illustration of a sinusoidally corrugated substrate with wavelength $\lambda$ and amplitide $\Delta$ and the associated solid-liquid and liquid-vapor interfaces. (b) Schematic illustration of planar and corrugated surfaces.}
\label{fig20}
\end{figure}

We now wish to consider a sinusoidally modulated surface, Eq.~(\ref{zwall}), on which a wall potential of the type of Eq.~(\ref{vtildel}) acts. We then expect that thermal equilibrium will be described by a sinusoidal modulation of the position of the interface $l(y)$ as well, i.e.
\begin{equation}
  l(y)=l_{\rm min}+A\sin(2\pi y/\lambda),
  \label{lysin}
\end{equation}
see Fig.~\ref{fig20}a; we thus assume that the modulation is in phase with the modulation of the wall position, and the task is to find the amplitude of this modulation, $A=A(\Delta,\lambda)$.  Physically, it is plausible that $A<\Delta$, so that the grooves  of the surface are to some extent filled with liquid and the ridges are correspondingly depleted. If $A=\Delta$, then the liquid-vapor interface would be just a uniform translation of the wall along the $z$-direction, and if $A=0$, the interface would be flat, i.e. the wall is not ``felt'' by the interface.

To compute $A(\Delta,\lambda)$, we need to make the proper choice of the wall potential $V(l,y)$ and it is tempting to assume:
\begin{equation}
  V(l,y)=V(l-z_{\rm wall}(y)),
  \label{vlyzwall}
\end{equation}
and use this in Eq.~(\ref{gammawvyl}). However, this assumption would neglect that a curved solid surface (when we assumed that the nearest-neighbor distance between surface atoms is always the same) contains more atoms (per unit area in the $xy$-plane) than a planar solid surface (as schematically sketched in Fig.~\ref{fig20}b). In the continuum limit, this is  in our case simply described by an enhancement with the Wenzel factor $r_{\rm w}$; hence we make the assumption that Eq.~(\ref{gammawvyl}) for the problem sketched in  Fig.~\ref{fig20}a needs to be replaced by (we now compute $\gamma_{wv}$ for a corrugated solid surface):
\begin{equation}
  \gamma_{wv}^{\rm corr}=\gamma_{lv}+\frac{1}{y_L}\int_{0}^{y_{L}} dy\left([1+(\pi\Delta/\lambda)^2][\gamma_{wl}(y)+V(l-\Delta\sin\frac{2\pi y}{\lambda})]
    +\frac{\gamma_{lv}}{2}\left(\frac{dl}{dy}\right)^2\right).
  \label{gammawvcorr1}
\end{equation}
when $A=\Delta$, $V(l-\Delta\sin\frac{2\pi y}{\lambda})=V(l_{\rm min})$,
$\frac{1}{2y_L}\int_{0}^{y_{L}}dy \left(\frac{dl}{dy}\right)^2=\left(\frac{\pi\Delta}{\lambda}\right)^2$, and then
\begin{equation}
  \gamma_{wv}^{\rm corr}=[1+(\pi\Delta/\lambda)^2][\gamma_{wl}+\gamma_{lv}+V(l_{\rm min})],
  \label{gammawvcorr2}
\end{equation}
and using Eqs.~(\ref{young}) and (\ref{costheta}) thus yields
\begin{equation}
  \gamma_{wv}^{\rm corr}=[1+(\pi\Delta/\lambda)^2]\gamma_{wl}+
  [1+(\pi\Delta/\lambda)^2]\gamma_{lv}\cos(\theta),
  \label{gammawvcorr3}
\end{equation}

Note that $[1+(\pi\Delta/\lambda)^2]\gamma_{wl}$ is simply nothing by $\gamma_{wl}^{\rm corr}$, the liquid-wall tension of the corrugated surface, so Eq.~(\ref{gammawvcorr3}) means
\begin{equation}
  \gamma_{wv}^{\rm corr}-\gamma_{wl}^{\rm corr}
  =[1+(\pi\Delta/\lambda)^2]\gamma_{lv}\cos(\theta)=\gamma_{lv}\cos(\theta_{\rm av}),
  \label{gammawvcorr4}
\end{equation}
where in the last step Eq.~(\ref{costhetaav}) was used. Thus the Ansatz Eq.~(\ref{gammawvcorr1}) does reproduce the Wenzel relation,  Eq.~(\ref{costhetaav}), in the limit when $A=1$. This condition is a necessary consistency condition when we consider a macroscopic corrugation ($\Delta\gg l_{\rm min}$, but nevertheless
$\Delta/\lambda\ll 1$ and droplet radius $R\gg \lambda$ so that only  the average effect of the corrugation matters). 

Still, the assumption Eq.~(\ref{gammawvcorr1}) is a special model assumption, which neglects other effects such as a change of the local potential $V(l-z_{\rm wall}(y))$ due to the local curvature of the corrugated wall. Curvature corrections to wall tensions have been considered for spheres and cylinders (e.g.~\cite{evans03,parry06}). We neglect them here for two reasons: (i) the average curvature of the sinusoidal corrugated surface is zero, so terms inverse in the first power of the radius of curvature should make no contribution, (ii) a correct description of the curvature effects requires a nonlocal theory for the interface potential~\cite{parry06}, which is beyond the scope of the present qualitative considerations.

\begin{figure} 
\includegraphics[scale=0.5]{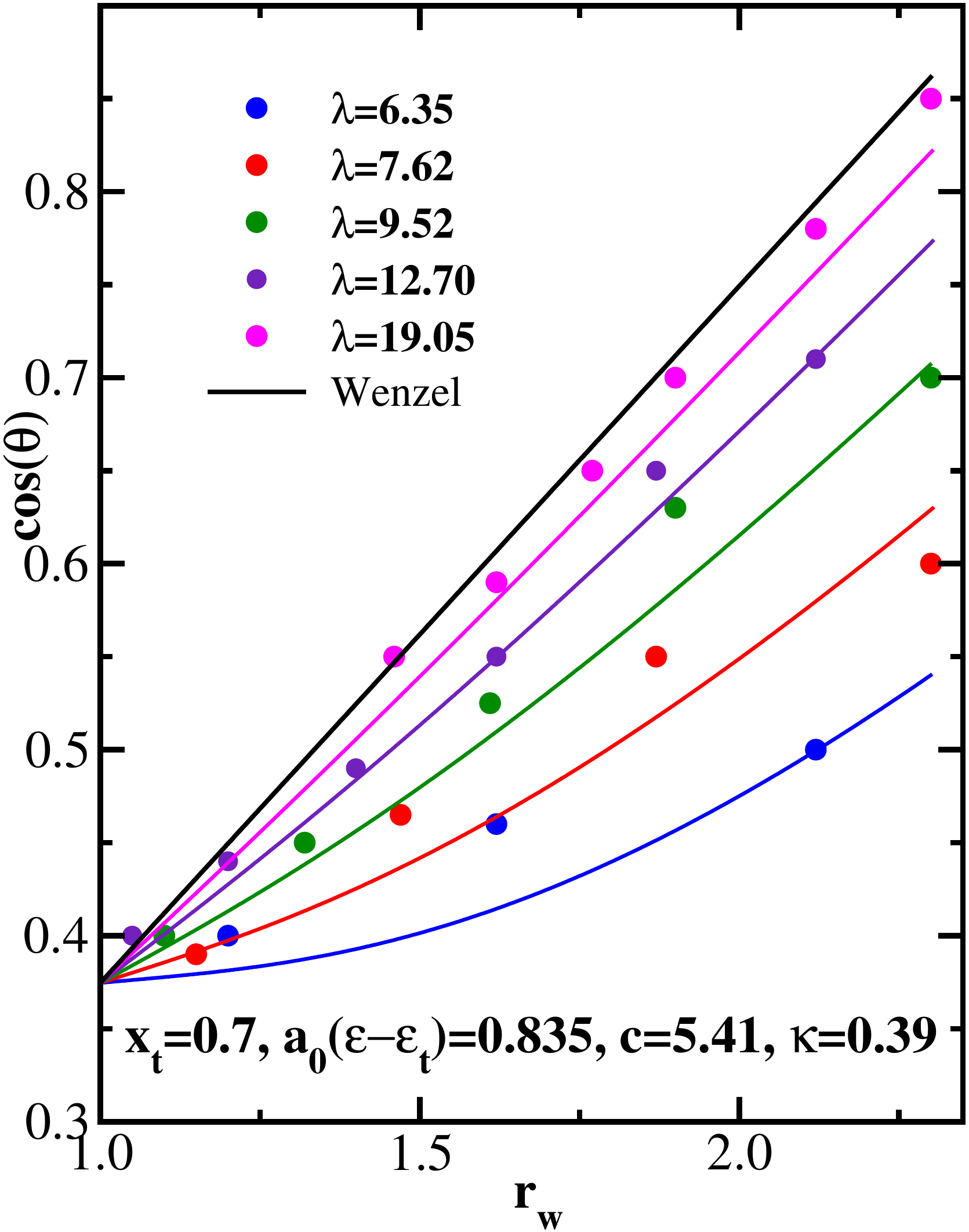} 
\caption{Theoretical (lines) and simulation~\cite{grzelak10} (symbols) results for the cosine of the contact angle as a function of the Wenzel parameter $r_{\rm w}$ for several values of the substrate wavelength $\lambda$, as indicated.} 
\label{fig21} 
\end{figure}

To find the solution of Eq.~(\ref{gammawvcorr1}), we have to solve the Euler-Lagrange equation for the free energy functional for $l(y)$ or, equivalently, for $\delta l(y)=l(y)-l_{\rm min}$:
\begin{equation}
  \left[1+\left(\frac{\pi\Delta}{\lambda}\right)^2\right]\frac{\partial \tilde{V}}{\partial (\delta l)}=
    \frac{d^2(\delta l)}{d y^2},
  \label{euler1}
\end{equation}
which yields, using Eq.~(\ref{vtildel}) and expanding $\tilde{V}$ linearly in
$\delta l-\Delta\sin\frac{2\pi y}{\lambda}$:
\begin{equation}
 \left[1+\left(\frac{\pi\Delta}{\lambda}\right)^2\right]\kappa^2 x(-a_0\delta\epsilon-4bx+9cx^2)
  (\delta l-\Delta\sin\frac{2\pi y}{\lambda})=
    \frac{d^2(\delta l)}{d y^2}.
  \label{euler2}
\end{equation}

Introducing the abbreviation
\begin{equation}
  C(\epsilon,x)=\left[1+\left(\frac{\pi\Delta}{\lambda}\right)^2\right](-a_0\delta\epsilon-4bx+9cx^2)
  \label{cepsilonx}
\end{equation}
and using Eq.~(\ref{lysin}) the differential equation~(\ref{euler2}) is solved by:
\begin{equation}
  A(\Delta,\lambda)=\Delta/\left[1+\left(\frac{2\pi}{\kappa\lambda}\right)^2/(xC(\epsilon,x)) \right].
  \label{adeltalambda}
\end{equation}

This solution now needs to be used in Eq.~(\ref{gammawvcorr1}), to compute the resulting shape of
$\gamma_{wv}^{\rm corr}$. The term $V(l(y)-\Delta\sin\frac{2\pi y}{\lambda})$ is again expanded in a power series in terms of  $(\delta l(y)-\Delta\sin\frac{2\pi y}{\lambda})$ (i.e. around the solution $l_{\rm  min}$ of the case without corrugation) for this purpose. As expected, the linear term vanishes when the integral over $y$ is performed from $y=0$ to $y=\lambda$. So the change resulting from this term arises only from the quadratic term of the Taylor expansion, yielding a correction of order $\Delta^2$. Also the term $(dl/dy)^2$  
yields a correction of order $\Delta^2$.

We thus conclude that the two corrections $\Delta H^{(1)}$ and $\Delta H^{(2)}$ to the wall tension of the vapor above the corrugated surface hence are:
\begin{equation}
  \Delta H^{(1)}/\gamma_{lv}=\left[1+\left(\frac{\pi\Delta}{\lambda}\right)^2\right]
  \frac{1}{\lambda}\int_{0}^{\lambda} dy
  \frac{1}{2}\left[\delta l(y)-\Delta\sin\frac{2\pi y}{\lambda} \right]^2\kappa^2
  (-a_0\delta\epsilon x-4bx^2+9cx^3)=\frac{1}{4}\kappa^2 xC(\epsilon,x)(A-\Delta)^2,
    \label{deltah1}
\end{equation}
  while the correction resulting from the term $(dl/dy)^2$ is 
\begin{equation}
  \Delta H^{(2)}/\gamma_{lv}=\frac{1}{4}A^2\left(\frac{2\pi}{\lambda}\right)^2=
  \left(\frac{\pi\Delta}{\lambda}\right)^2/
  \left[1+\left(\frac{2\pi}{\kappa\lambda}\right)^2/(xC(\epsilon,x)) \right].
    \label{deltah2}
\end{equation}
After a little algebra one finds:
\begin{equation}
  \Delta H^{(1)}+\Delta H^{(2)}=\gamma_{lv}\left(\frac{\pi\Delta}{\lambda}\right)^2/
\left[1+\left(\frac{2\pi}{\kappa\lambda}\right)^2/(xC(\epsilon,x)) \right],
    \label{deltah12}
\end{equation}
and this implies that in the limit where $\kappa\lambda\rightarrow\infty$ a correction
$\gamma_{lv}\left(\frac{\pi\Delta}{\lambda}\right)^2$ remains, and Eq.~(\ref{deltah12}) reduces to
Eq.~(\ref{gammawvcorr2}), as it should on physical grounds.

In order to discuss the corrections to Wenzel's result more explicitly, we note that $C(\epsilon,x)$ given by Eq.~(\ref{cepsilonx}) can be expressed in terms of the constants $x_t$, $a_0(\epsilon-\epsilon_t)$, and $c$ as follows:
\begin{equation}
  C(\epsilon,x)=\left[1+\left(\frac{\pi\Delta}{\lambda}\right)^2\right]
  c\left\{\frac{2x_{t}^{2}}{3}+2a_0(\epsilon-\epsilon_t)
  +\frac{4}{3}x_{t}^{2}\sqrt{1+3\frac{a_0(\epsilon-\epsilon_t)}{cx_{t}^{2}} }   \right\}
\label{cepsilonx2}
\end{equation}

We note that for $\epsilon=\epsilon_t$ the curly bracket simply yields $2x_{t}^{2}$, and hence Eq.~(\ref{deltah12}) then yields for the correction to the vapor-wall surface  tension the result
\begin{equation}
  \Delta H^{(1)}+\Delta H^{(2)}=\gamma_{lv}\left(\frac{\pi\Delta}{\lambda}\right)^2/
\left[1+\frac{(2\pi/(\kappa\lambda))^2}{[1+(\pi\Delta/\lambda)^2]2cx_{t}^{3}} \right],
    \label{deltah12two}
\end{equation}
and hence (note that $V(l_{\rm min})=0$ for $\epsilon=\epsilon_t$)
\begin{equation}
  \gamma_{wv}^{\rm corr}=\gamma_{lv}\left\{1+\frac{(\pi\Delta/\lambda)^2}
    {1+\frac{(2\pi/(\kappa\lambda))^2}{[1+(\pi\Delta/\lambda)^2]2cx_{t}^{3}}}   \right\}
  +\gamma_{wl}\left[1+(\pi\Delta/\lambda)^2\right],
  \label{gammawvcorr5}
\end{equation}
  and using that $\gamma_{lv}+\gamma_{wl}=\gamma_{wv}$ at the wetting transition of a flat wall, one obtains
\begin{equation}
  \gamma_{wv}^{\rm corr}=\gamma_{wv}\left[1+(\pi\Delta/\lambda)^2\right]-
  \gamma_{lv}\left(\frac{\pi\Delta}{\lambda}\right)^2
  \frac{1}{\left[1+(\pi\Delta/\lambda)^2\right]2cx_{t}^{3}(\kappa\lambda/(2\pi))^2+1}.
  \label{gammawvcorr6}
\end{equation}
The Wenzel result would be $\gamma_{wv}^{\rm corr}=\gamma_{wv}[1+(\pi\Delta/\lambda)^2]$, and
so we see that the theory predicts a correction term $\Delta \gamma_{wv}^{\rm corr}$, which for small values of
$(2\pi/(\kappa\lambda))$ simply becomes
\begin{equation}
  \Delta\gamma_{wv}^{\rm corr}=-\gamma_{lv}\left(\frac{\pi\Delta}{\lambda} \right)^2
  \left(\frac{2\pi}{\kappa\lambda} \right)^2/\left[1+\left(\frac{\pi\Delta}{\lambda}
    \right)^2 2cx_{t}^{3}\right],
  \label{deltagammawvcorr}
  \end{equation}
  which implies a shift of the wetting transition due to corrugation which scales proportionally to
  $(\kappa\lambda)^{-2}$, as well as $(\pi\Delta/\lambda)^2$.

  For $\epsilon\neq\epsilon_t$, it is instructive to cast the present result in the form:
  \begin{equation}
    (\gamma_{wv}^{\rm corr}-\gamma_{wl}^{\rm corr})/\gamma_{lv}=\cos(\theta_{\rm eff})=
    r_{\rm w}\cos\theta-\left(\frac{\pi\Delta}{\lambda} \right)^2
    \frac{1}
    {1+\left(\frac{\kappa\lambda}{2\pi} \right)^2xC(\epsilon,x)}.
    \label{rwcorrection}
 \end{equation}
 As it must be, the deviations from Wenzel's law vanish when $\kappa\lambda\rightarrow\infty$, i.e. for macroscopic corrugations.

 We now apply this theory to analyze the simulation data~\cite{grzelak10} for the dependence of the contact angle on the roughness of a sinusoidal substrate. The simulation study~\cite{grzelak10} used a microscopic model where the particles of monatomic fluid interact with each other and with the substrate particles via truncated and shifted LJ potential. With the parameters chosen in this work~\cite{grzelak10}, the cosine of the contact angle on a flat substrate takes the value $\cos\theta=0.375$. We treat $x_t$, $\kappa$, $c$, and $a_0(\epsilon-\epsilon_t)$ as adjustable parameters and perform the best fit of the simulation data given in the upper panel of Fig.~9 of Ref.~\cite{grzelak10} to Eq.~(\ref{rwcorrection}), which yields the following values:        
$x_t=0.70$, $\kappa=0.39$, $c=5.41$, and $a_0(\epsilon-\epsilon_t)=0.835$. The corresponding results for $\cos\theta$ as a function of $r_{\rm w}$ are shown in Fig.~\ref{fig21} together with the simulation data~\cite{grzelak10}, and the agreement is satisfactory. Similar to 2D-DFT results of Section~\ref{section3} the Wenzel behavior is gradually approached as the wavelength $\lambda$ (in units of the particle diameter) exceeds the value of 20. While the qualitative agreement between our
model and the simulation results~\cite{grzelak10} certainly is encouraging, we note
that the simulations did consider the case 
where $\Delta$  and $\lambda$ are of the same order, while the theory considers
the limit $\Delta/\lambda\ll 1$ only. But it is also encouraging, that the
filling or partial filling of the grooves did not seem to create any problems for the
simulations, however.


\begin{thebibliography}{10}

\bibitem{degennes04}
P.~G. de~Gennes, F.~Brochard-Wyart, and D.~Qu{\'e}r{\'e}, {\em Capillary and
  Wetting Phenomena -- Drops, Bubbles, Pearls, Waves}
\newblock (Springer: New York  2004).

\bibitem{butt03}
H.~J. Butt, K.~Graf, and M.~Kappl, {\em Physics and Chemistry of Interfaces}
\newblock (Wiley-VCH: Weinheim  2003).

\bibitem{ondarcuhu13}
T.~Ondar\c{c}uhu and J.~P. Aime{\'e}, {\em Nanoscale Liquid Interfaces:
  Wetting, Patterning and Force Microscopy at the Molecular Scale}
\newblock (Pan Stanford Publishing Pte Ltf: Stanford  2013).

\bibitem{bonn09}
D.~Bonn, J.~Eggers, J.~Indekeu, J.~Meunier, and E.~Rolley,
\newblock Rev.~Mod.~Phys.  {\bf 81},  739   (2009).

\bibitem{erbil14}
H.~Y. Erbil,
\newblock Surface Sci. Rep.  {\bf 69},  325   (2014).

\bibitem{young1805}
T.~Young,
\newblock Phil. Trans. Royal Soc. (London)  {\bf 95},  65   (1805).

\bibitem{gibbs61}
J.~W. Gibbs, {\em The Scientific Papers, Vol. 1}
\newblock (Dover Publ.: New York  1961).

\bibitem{rowlinson82}
J.~S. Rowlinson and B.~Widom, {\em Molecular Theory of Capillarity}
\newblock (Clarendon: Oxford  1982).

\bibitem{amirfazli04}
A.~Amirfazli and A.~W. Neumann,
\newblock Adv. Colloid Interface Sci.  {\bf 110},  121   (2004).

\bibitem{schimmele07}
L.~Schimmele, M.~Napiorkowski, and S.~Dietrich,
\newblock J.~Chem.~Phys.  {\bf 127},  164715   (2007).

\bibitem{wenzel36}
R.~N. Wenzel,
\newblock Ind. Eng. Chem.  {\bf 28},  988   (1936).

\bibitem{binder11b}
K.~Binder and W.~Kob, {\em Glassy Materials and Disordered Solids: An
  Introduction to Their Statistical Mechanics}
\newblock (World Scientific: Singapore  2011).

\bibitem{li90}
H.~Li and M.~Kardar,
\newblock Phys.~Rev.~B  {\bf 42},  6546   (1990).

\bibitem{johnson64}
R.~E. Johnson and R.~H. Dettre,
\newblock {\em Contact Angle, Wettability, and Adhesion}, vol~43 of {\em
  Advances in Chemistry}, pp 112-135,
\newblock Amer. Chem. Soc., F. M. Fowkes, Ed., Washington, DC, (1964).

\bibitem{dettre64}
R.~H. Dettre and R.~E. Johnson,
\newblock {\em Contact Angle, Wettability, and Adhesion}, vol~43 of {\em
  Advances in Chemistry}, pp 136-144,
\newblock Amer. Chem. Soc., F. M. Fowkes, Ed., Washington, DC, (1964).

\bibitem{swain98}
P.~S. Swain and R.~Lipowsky,
\newblock Langmuir  {\bf 14},  6772   (1998).

\bibitem{wolansky98}
G.~Wolansky and A.~Marmur,
\newblock Langmuir  {\bf 14},  5292   (1998).

\bibitem{quere08}
D.~Qu\'{e}r\'{e},
\newblock Annu. Rev. Mater. Res.  {\bf 38},  71   (2008).

\bibitem{velarde11}
Ed. M.~G.~Velarde, {\em Discussion and Debate: Wetting and Spreading Science -
  Quo Vadis?}
\newblock (EDP Sciences: Les Ulis  2011).

\bibitem{herminghaus08}
S.~Herminghaus, M.~Brinkmann, and R.~Seemann,
\newblock Ann. Rev. Mater. Res.  {\bf 38},  101   (2008).

\bibitem{hofmann10}
T.~Hofmann, M.~Tasinkevych, A.~Checco, E.~Dobisz, S.~Dietrich, and B.~M. Ocko,
\newblock Phys.~Rev.~Lett.  {\bf 104},  106102   (2010).

\bibitem{butt13}
H.~J. Butt, C.~Semprebon, P.~Papadopoulos, D.~Vollmer, M.~Brinkmann, and
  M.~Ciccotti,
\newblock Soft Matter  {\bf 9},  418   (2013).

\bibitem{xu14}
X.~M. Xu, G.~Vereecke, C.~Chen, G.~Pourtois, S.~Armini, N.~Verellen, W.~K.
  Tsai, D.~W. Kim, E.~Lee, C.~Y. Lin, P.~Van Dorpe, H.~Struyf, F.~Holsteyns,
  V.~Moshchalkov, J.~Indekeu, and S.~De Gendt,
\newblock ACS Nano  {\bf 8},  885   (2014).

\bibitem{tretyakov16}
N.~Tretyakov, P.~Papadopoulos, D.~Vollmer, H.-J. Butt, B.~D{\"u}nweg, and K.~C.
  Daoulas,
\newblock J.~Chem.~Phys.  {\bf 145},  134703   (2016).

\bibitem{rascon00}
C.~Rascon and A.~O. Parry,
\newblock Nature  {\bf 407},  986   (2000).

\bibitem{mickel11}
W.~Mickel, L.~Joly, and T.~Biben,
\newblock J.~Chem.~Phys.  {\bf 134},  094105   (2011).

\bibitem{berim11}
G.~O. Berim and E.~Ruckenstein,
\newblock J. Coll. Interface Sci  {\bf 359},  304   (2011).

\bibitem{malijevsky14}
A.~Malijevsky,
\newblock J.~Chem.~Phys.  {\bf 141},  184703   (2014).

\bibitem{malijevsky14b}
A.~Malijevsky and A.~O. Parry,
\newblock J.~Phys.~Cond.~Matt.  {\bf 26},  355003   (2014).

\bibitem{svoboda15}
M.~Svoboda, A.~Malijevsky, and M.~Lisal,
\newblock J.~Chem.~Phys.  {\bf 143},  104701   (2015).

\bibitem{zhou18}
S.~Q. Zhou,
\newblock J. Stat. Phys.  {\bf 170},  979   (2018).

\bibitem{malijevsky19}
A.~Malijevsky,
\newblock Phys.~Rev.~E  {\bf 99},  040801(R)   (2019).

\bibitem{daub10}
C.~D. Daub, J.~H. Wang, S.~Kudesia, D.~Bratko, and A.~Luzar,
\newblock Faraday Disc.  {\bf 146},  67   (2010).

\bibitem{grzelak10}
E.~M. Grzelak and J.~R. Errington,
\newblock Langmuir  {\bf 26},  13297   (2010).

\bibitem{leroy11}
F.~Leroy and F.~Muller-Plathe,
\newblock Langmuir  {\bf 27},  637   (2011).

\bibitem{kumar13}
V.~Kumar and J.~R. Errington,
\newblock Langmuir  {\bf 29},  11815   (2013).

\bibitem{chialvo13}
A.~A. Chialvo, L.~Vlcek, and P.~T. Cummings,
\newblock J.~Phys.~Chem.~C  {\bf 117},  23875   (2013).

\bibitem{tretyakov13}
N.~Tretyakov and M.~M{\"u}ller,
\newblock Soft Matter  {\bf 9},  3613   (2013).

\bibitem{fortini13}
A.~Fortini and M.~Schmidt,
\newblock Soft Matter  {\bf 15},  3994   (2013).

\bibitem{ambrosia18}
M.~S. Ambrosia and M.~Y. Ha,
\newblock Computers and Fluids  {\bf 163},  1   (2018).

\bibitem{gao07}
L.~C. Gao and T.~J. McCarthy,
\newblock Langmuir  {\bf 23},  3762   (2007).

\bibitem{marmur11}
A.~Marmur,
\newblock Eur. Phys. J. Special Topics  {\bf 197},  193   (2011).

\bibitem{egorov18d}
S.~K. Das, S.~A. Egorov, P.~Virnau, D.~Winter, and K.~Binder,
\newblock J.~Phys.~Cond.~Matt.  {\bf 30},  255001   (2018).

\bibitem{tolman49}
R.~C. Tolman,
\newblock J.~Chem.~Phys.  {\bf 17},  333   (1949).

\bibitem{troster18}
A.~Tr{\"o}ster, F.~Schmitz, P.~Virnau, and K.~Binder,
\newblock J.~Phys.~Chem.~B  {\bf 122},  3407   (2018).

\bibitem{henderson05}
J.~R. Henderson,
\newblock Mol. Phys.  {\bf 103},  2839   (2005).

\bibitem{rodriguezrivas15}
A.~Rodriguez-Rivas, J.~Galvan, and J.~M. Romero-Enrique,
\newblock J. Phys. Cond. Matt.  {\bf 27},  035101   (2015).

\bibitem{yates74}
R.~C. Yates, {\em Curves and their Properties}
\newblock (National Council of Teachers of Mathematics:  1974).

\bibitem{farouki90}
R.~T. Farouki and C.~A. Neff,
\newblock Comp. Aided Geom. Design  {\bf 7},  83   (1990).

\bibitem{alejandre07}
J.~Alejandre, F.~Bresme, M.~Gonzalez-Melchor, and F.~del Rio,
\newblock J.~Chem.~Phys.  {\bf 126},  224511   (2007).

\bibitem{lekkerkerker11}
H.~N.~W. Lekkerkerker and R.~Tuinier, {\em Colloids and the Depletion
  Interaction}
\newblock (Springer: Dordrecht  2011).

\bibitem{sitta16}
C.~E. Sitta, F.~Smallenburg, R.~Wittkowski, and H.~L{\"o}wen,
\newblock J.~Chem.~Phys.  {\bf 145},  204508   (2016).

\bibitem{kim12}
E.~Y. Kim, S.~C. Kim, and B.~S. Seong,
\newblock J.~Phys.~Chem.~B  {\bf 116},  3180   (2012).

\bibitem{swain98b}
P.~S. Swain and A.~O. Parry,
\newblock Eur. Phys. J. B  {\bf 4},  459   (1998).

\bibitem{rascon00b}
C.~Rascon and A.~O. Parry,
\newblock J.~Phys.~Cond.~Matt.  {\bf 12},  A369   (2000).

\bibitem{rejmer07}
K.~Rejmer,
\newblock Physica A  {\bf 373},  58   (2007).

\bibitem{dietrich88}
S.~Dietrich,
\newblock {\em Wetting Phenomena}, vol~12 of {\em Phase Transitions and
  Critical Phenomena, C. Domb and J. L. Lebowitz, eds.}, pp 1-218,
\newblock Academic Press, New York, (1988).

\bibitem{parry04}
A.~O. Parry, J.~M. Romero-Enrique, and A.~Lazaridis,
\newblock Phys.~Rev.~Lett.  {\bf 93},  086104   (2004).

\bibitem{evans03}
R.~Evans, R.~Roth, and P.~Bryk,
\newblock Europhys.~Lett.  {\bf 62},  815   (2003).

\bibitem{parry06}
A.~O. Parry, C.~Rascon, and L.~Morgan,
\newblock J.~Chem.~Phys.  {\bf 124},  151101   (2006).

\end{thebibliography}

\end{document}